\pdfoutput=1
\documentclass[1p,10pt]{elsarticle}
\journal{Combustion and Flame}

\biboptions{sort&compress, square, comma}
\usepackage[linkcolor=blue, citecolor=blue, colorlinks=true]{hyperref}

\usepackage[version=3]{mhchem} 
\usepackage{amsmath,amssymb}

\usepackage{booktabs,multicol}

\usepackage{graphicx}
\usepackage{caption}
\usepackage{subcaption}

\usepackage[english]{babel}
\usepackage{textcomp}

\usepackage{siunitx}
\DeclareSIUnit\atm{atm}

\begin{document}
\begin{frontmatter}

\title{Mechanism reduction for multicomponent surrogates: a case study using toluene reference fuels}

%

\author[osu]{Kyle~E.\ Niemeyer\corref{cor1}}
\ead{Kyle.Niemeyer@oregonstate.edu}

\author[uconn]{Chih-Jen Sung}
\ead{cjsung@engr.uconn.edu}

\address[osu]{School of Mechanical, Industrial, and Manufacturing Engineering\\
	Oregon State University, Corvallis, OR 97331, USA}
\address[uconn]{Department of Mechanical Engineering\\
	University of Connecticut, Storrs, CT 06269, USA}

\cortext[cor1]{Corresponding author}

\begin{abstract}
Strategies and recommendations for performing skeletal reductions of multicomponent surrogate fuels are presented, through the generation and validation of skeletal mechanisms for a three-component toluene reference fuel.
Using the directed relation graph with error propagation and sensitivity analysis method followed by a further unimportant reaction elimination stage, skeletal mechanisms valid over comprehensive and high-temperature ranges of conditions were developed at varying levels of detail.
These skeletal mechanisms were generated based on autoignition simulations, and validation using ignition delay predictions showed good agreement with the detailed mechanism in the target range of conditions.
When validated using phenomena other than autoignition, such as perfectly stirred reactor and laminar flame propagation, tight error control or more restrictions on the reduction during the sensitivity analysis stage were needed to ensure good agreement.
In addition, tight error limits were needed for close prediction of ignition delay when varying the mixture composition away from that used for the reduction.
In homogeneous compression-ignition engine simulations, the skeletal mechanisms closely matched the point of ignition and accurately predicted species profiles for lean to stoichiometric conditions.
Furthermore, the efficacy of generating a multicomponent skeletal mechanism was compared to combining skeletal mechanisms produced separately for neat fuel components; using the same error limits, the latter resulted in a larger skeletal mechanism size that also lacked important cross reactions between fuel components.
Based on the present results, general guidelines for reducing detailed mechanisms for multicomponent fuels are discussed.
\end{abstract}


\end{frontmatter}

\section{Introduction}
\label{S:intro}

Liquid transportation fuels such as gasoline, diesel, and jet fuel consist of a near-continuous spectrum of constituent hydrocarbons, including linear and branched alkanes, aromatics, cycloalkanes, and alkenes, with a specific makeup that varies depending on the feedstock source and refinery processes used.
This complexity and varying makeup pose challenges to computational modeling, so in order to represent real fuels the combustion community embraced the concept of surrogate fuels.
Since Edgar~\cite{Edgar:1927fq} first proposed using n-heptane and iso-octane as reference fuels from which to measure the knocking characteristics of gasoline---resulting in the octane number---various neat fuels and combinations of fuels have been used to represent complex liquid transportation fuels.
For example, Sturgis and coworkers~\cite{Sturgis:1954uq,Pahnke:1954er} used diisobutylene and benzene to represent olefins and aromatics, respectively, present in gasoline for antiknock studies of gasoline-like hydrocarbon mixtures.
The concept of grouping chemical species based on class also occurs in other areas such as petroleum processing, where the PONA analysis separates all hydrocarbons into paraffins, olefins, naphthenes, and aromatics~\cite{Universal-Oil-Products-Co.:1940}.
For the past five decades, the petroleum processing and chemical kinetics fields used various lumping techniques~\cite{Wei:1969kt,Coxson:1987uj,Bailey:2001gp,Lu:2008bi,Pepiot:2008kq} that group together chemical species of the same class (or with similar characteristics) and consider using a single representative---reducing the complexity of the kinetic system considerably.

As detailed reaction mechanisms for the larger hydrocarbons present in liquid transportation fuels were developed---such as the classic n-heptane mechanism of Curran et al.~\cite{Curran:1998}---and, simultaneously, the need for advanced engine concepts demanded more accurate representation of transportation fuels, significant research efforts focused on creating better surrogate models for these fuels.
Surrogates attempt to emulate the physical and chemical properties of complex real fuels by using specific mixtures of neat components, typically chosen to represent major classes of hydrocarbons present in the target real fuel.
For example, as discussed above engine modelers traditionally represented gasoline using a binary mixture of the so-called primary reference fuels (PRFs): n-heptane and iso-octane, which represent linear and branched alkanes, respectively.
More recently, additional components such as toluene---representing aromatics---have been considered to better match the reactivity of gasoline, forming the three-component toluene reference fuels (TRFs)~\cite{Gauthier:2004,Andrae:2007,Chaos:2007,Andrae:2008,Liang:2009a,Morgan:2010,Kumar:2010js,Hartmann:2011ku,Knop:2013gt,Kukkadapu:2013ko,Rapp:2013ev}.

A large number of surrogates for gasoline, diesel, and jet fuels have been proposed, as discussed in a number of recent surveys~\cite{Pitz:2007,Farrell:2007,Pitz:2011iv,Colket:2007,Colket:2008}, with a trend towards increasing numbers of components and complexity over time.
As mentioned earlier, binary PRF and ternary TRF mixtures have been used to represent gasoline; recently proposed four-component surrogates supplemented the TRF fuels with an additional component to represent olefins (alkenes), such as 1-hexene~\cite{Mehl:2011cn}, 1-pentene~\cite{Lenhert:2009}, and 2-pentene~\cite{Mehl:2011jn}.
Naik et al.~\cite{Naik:2005} proposed a gasoline surrogate with a component for each of the major constituent hydrocarbon classes, adding 1-pentene and methylcyclohexane to the three TRF components.

Similar to the progress in gasoline, a number of surrogates for diesel have also recently been proposed.
Some studies used neat fuels such as n-heptane and n-tetradecane~\cite{Puduppakkam:2011} to represent diesel, although---as with gasoline---recently more complex mixtures of multiple fuels have been proposed as diesel surrogates: binary mixtures of n-decane and 1-methylnaphthalene (the IDEA surrogate)~\cite{Hentschel:1994}, and n-heptane and toluene~\cite{Hernandez:2008bv}; a ternary mixture of n-propylcyclohexane, n-butylbenzene, and 2,2,4,4,6,8,8-heptamethylnonane~\cite{Mathieu:2009he}; and a five-component mixture of n-hexadecane, iso-octane, 1-methylnaphthalene, n-propylcyclohexane, and n-propylbenzene~\cite{Mati:2007hh}.
Mueller et al.~\cite{Mueller:2012gx} recently presented a methodology for constructing surrogates to closely match the composition, ignition quality, and volatility of diesel fuels, proposing an eight-component diesel surrogate.
Refer to Pitz and Mueller~\cite{Pitz:2011iv} or Krishnasamy et al.~\cite{Krishnasamy:2013bx} for more detailed surveys of diesel surrogates.

Biodiesels, produced from vegetable oils and animal fat, consist primarily of long-chain methyl esters.
Originally, methyl butanoate was used as a surrogate fuel, but it poorly represents the kinetics of real biodiesels~\cite{Herbinet:2008iw}.
Surrogates for biodiesel fuels proposed more recently include methyl decanoate~\cite{Herbinet:2008iw}; binary mixtures of methyl butanoate and n-heptane~\cite{Liu:2013em}; and a three-component mixture of methyl decanoate, methyl-9-decenoate, and n-heptane~\cite{Herbinet:2010}.
Proposed surrogates for kerosene-type jet fuels include mixtures of n-decane and 1,2,4-trimethylbenzene~\cite{Honnet:2009kl}; n-decane, iso-octane, and toluene~\cite{Dooley:2010js}; a four-component mixture of n-dodecane, iso-octane, 1,3,5-trimethylbenzene, n-propylbenzene~\cite{Dooley:2012bp}; and even a five-component mixture of methylcyclohexane, toluene, benzene, iso-octane, and n-dodecane~\cite{Violi:2002hl}.

Despite some success in developing surrogate fuels that emulate real fuel properties, unfortunately detailed reaction mechanisms for these surrogates typically consist of thousands of species and many more reactions.
For example, detailed mechanisms for three-component surrogates of gasoline and biodiesel contain \num{1389} species and \num{5935} reactions~\cite{Mehl:2011cn} and \num{3329} species and \num{10806} reactions~\cite{Herbinet:2010}, respectively.
In the computational simulations where engine modelers want to use these kinetic mechanisms, the cost of detailed chemistry scales cubically with the number of species in the worst case, due to the Jacobian matrix factorization required in implicit integration algorithms~\cite{Lu:2009gh}.
Detailed mechanisms of such large sizes are time-consuming even in homogeneous or one-dimensional simulations, and pose prohibitive computational challenges to implementation in multidimensional simulations.
Therefore, using these mechanisms requires mechanism reduction and simplification.

While implementing such large, multicomponent detailed mechanisms poses significant challenges to combustion simulations, generating complex reaction mechanisms with potentially thousands of species and reactions presents a difficult task in its own right.
In the past, most models were constructed based on the application of reaction classes, and recently several large detailed mechanisms were produced in this manner (e.g., \cite{Westbrook:2009,Sarathy:2011}).
However, for large hydrocarbons and particularly for multicomponent surrogate mechanisms, this process becomes impractical as the fuel molecule size increases, so automated mechanism generation methods are an active area of research~\cite{Blurock:1995tj,Ranzi:1995,Come:1996,Warth:2000bx,Ranzi:2005fy,Moreac:2006ej,VanGeem:2006fv,Hansen:2013fe}.
Interested readers should see reviews by Tomlin et al.~\cite{Tomlin:1997}, Pierucci and Ranzi~\cite{Pierucci:2008dl}, and Blurock et al.~\cite{Blurock:2013vc} for more discussion on this topic.

Classical approaches to mechanism reduction rely on chemical kinetics understanding and expertise, but in recent years a number of automated, theory-based methods were developed in response to the ever-increasing size and complexity of detailed mechanisms.
Although an extensive survey of such methods is beyond the scope of the current work---see Lu and Law~\cite{Lu:2009gh} for a recent review---one class of methods in particular recently received significant development: the directed relation graph (DRG) approach of Lu and Law~\cite{Lu:2005ce,Lu:2006bb,Lu:2006gi}.
Similar to the earlier graphical representation of reaction pathways developed by Bendtsen et al.~\cite{Bendtsen:2001vh}, the DRG method quantifies the importance of species by mapping interspecies coupling to a weighted directed graph and trimming unimportant connections in the resulting graph using a cutoff threshold.
Then, it performs a graph search initiated at selected target species to find connected species---the reachable set of species makes up the skeletal mechanism.
Variants of the original DRG method include DRG with error propagation (DRGEP)~\cite{Pepiot-Desjardins:2008,Niemeyer:2011fe}, which considers the geometric propagation of dependence down graph pathways such that more distant species are less important, and path-flux analysis~\cite{Sun:2010jf}, which considers separately direct and indirect (through a single intermediate species) contributions to species production and consumption rates.
In addition, DRG and DRGEP have been combined with a sensitivity analysis of remaining species to produce a more compact skeletal mechanism than possible with the original methods alone~\cite{Sankaran:2007fs,Zheng:2007gd,Lu:2008bi,Niemeyer:2010bt}.
More recently, Luo et al.~\cite{Luo:2010cu} updated the DRG interaction coefficient to better represent interspecies dependence in the presence of many isomers.

However, most studies applying the above methods---and other reduction approaches---used detailed mechanisms for neat (single component) fuels.
In particular, efforts focused mainly on normal and branched alkanes such as n-heptane~\cite{Lu:2006bb,Lu:2008bi,Niemeyer:2010bt,Shi:2010a,Sun:2010jf,Bahlouli:2012kq}, iso-octane~\cite{Lu:2006bb,Pepiot-Desjardins:2008,Niemeyer:2010bt,Shi:2010}, and n-decane~\cite{Niemeyer:2010bt,Sun:2010jf}, with the exception of Luo et al.~\cite{Luo:2010cu,Luo:2012dp,Luo:2012cr}, who generated skeletal mechanisms for a three-component biodiesel surrogate (n-heptane, methyl decanoate, and methyl-9-decenoate)~\cite{Herbinet:2010} at various conditions using DRG-aided sensitivity analysis.
Luo et al.~\cite{Luo:2010cu} showed the performance of their high-temperature skeletal mechanism for some off-target conditions---i.e., at mixture compositions different than that used for the reduction---but they did not perform a detailed analysis of this or provide general guidelines for reducing multicomponent detailed mechanisms.
Since the oxidation of different fuel species in a multicomponent blend initiates at different conditions and occurs along different reaction pathways, the combustion characteristics (e.g., autoignition delay, laminar flame speed) of fuel blends are in general nonlinear functions of the fuel composition.
Interactions between fuel components induce added complexity compared to a neat fuel, causing a greater challenge for mechanism reduction.
Other efforts to generate compact skeletal mechanisms for multicomponent surrogates involved combining existing skeletal mechanisms for various components~\cite{Andrae:2008,Blanquart:2009iw,Ra:2011gh}.
However, it is unclear whether this approach will ensure the comprehensiveness of the resulting combined skeletal mechanisms by capturing nonlinear chemical interaction effects.
In addition, this straightforward merging may lead to an undesired larger size of the final skeletal mechanism.

For our mechanism reduction approach, we selected a multi-stage technique using the DRGEPSA method~\cite{Niemeyer:2010bt} followed by an unimportant reaction elimination stage~\cite{Lu:2008bi}.
We chose the primary reduction stage focusing on species removal because DRG-based methods with sensitivity analysis demonstrate effective reduction capabilities on large detailed mechanisms at a reasonable computational expense~\cite{Niemeyer:2010bt,Luo:2010cu,Shi:2010a,Luo:2012dp,Luo:2012cr}.
Similarly, while removing species simultaneously reduces the number of reactions, we added a secondary unimportant reaction elimination step to further decrease the size of the mechanism.
Additional reduction stages such as isomer lumping and time-scale reduction~\cite{Lu:2008bi} could also be applied for an objective of producing the smallest possible mechanism, although the steps we consider here offer the most substantial reduction in mechanism size.
Regarding isomer lumping, Luo et al.~\cite{Luo:2012cr} found that few isomers could be lumped together when considering low-temperature chemistry, so such a reduction step may only be useful for producing skeletal mechanisms relevant at high temperatures.
With these comments in mind, rather than simply generate a single multicomponent skeletal mechanism for a particular surrogate fuel, recognizing the large and growing number of surrogates---and the large and growing size of corresponding detailed mechanisms---the prime focus of the current work is to study the behavior of the skeletal reduction algorithms with multicomponent fuels, and provide recommendations for best practices in such applications using the above and related mechanism reduction methods.

Our objective in the current work was to establish guidelines and strategies for the reduction of detailed mechanisms for multicomponent fuels, by 
\begin{enumerate}
\item performing skeletal reductions of a detailed mechanism for a three-component TRF, commonly used to represent gasoline;
\item validating the resulting skeletal mechanisms for performance in the target range of conditions for autoignition;
\item investigating the performance in other combustion applications (e.g., PSR and laminar flame propagation); 
\item comparing the ability of the skeletal mechanisms to predict species evolution in homogeneous charge compression ignition (HCCI) simulations;
\item studying the performance when used outside the target range---in other words, at other surrogate mixture compositions; and 
\item comparing the efficacy of this multicomponent reduction approach to that of combining separate skeletal mechanisms for neat fuels, in terms of comprehensiveness and mechanism size.
\end{enumerate}
In doing so, we showed that, in general, tight error control may be needed for wide-ranging accuracy.

\section{Methodology}
\label{S:method}

\subsection{DRGEP}
\label{S:DRGEP}

In the current work we used the DRGEPSA method of Niemeyer et al.~\cite{Niemeyer:2010bt}, which is based on the DRGEP method of Pepiot-Desjardins and Pitsch~\cite{Pepiot-Desjardins:2008} followed by a sensitivity analysis of certain remaining species.
First, the DRGEP method is described here.
The dependence of a species \emph{A} that requires another species \emph{B} in the mechanism (in order to accurately calculate its production) is expressed with the direct interaction coefficient:
\begin{equation}
\label{E:dic}
	r_{AB} = \frac{ \bigl | \sum_{i = 1}^{N_R} \nu_{A, i} \Omega_{i} \delta_{B}^{i} \bigr | } {\max \left( P_A, C_A \right)  },
\end{equation}
where $P_A$ and $C_A$ are production and consumption terms given by
\begin{align}
	P_A &= \sum_{i=1}^{N_R} \max \left( 0, \nu_{A,i} \Omega_i \right) \text{ and}		\label{E:PA} \\
	C_A &= \sum_{i=1}^{N_R} \max \left( 0, - \nu_{A,i} \Omega_i \right),		\label{E:CA}
\end{align}
respectively. In the above equations,
\begin{equation}
\label{E:deltaBi}
	\delta_{B}^{i} =
	\begin{cases}
		1 &\text{if reaction \emph{i} involves species \emph{B},} \\
		0 &\text{otherwise,}
	\end{cases}
\end{equation}
\emph{A} and \emph{B} represent the species of interest (with dependency in the $A \rightarrow B$ direction meaning that \emph{A} depends on \emph{B}), \emph{i} the \emph{i}th reaction, $\nu_{A, i}$ the overall stoichiometric coefficient of species \emph{A} in the \emph{i}th reaction ($\nu_{A, i} = \nu_{A, i}^{\prime \prime} - \nu_{A, i}^{\prime}$, where $\nu^{\prime\prime}$ and $\nu^{\prime}$ are the product and reactant coefficients, respectively), $\Omega_i$ the overall reaction rate of the \emph{i}th reaction, and $N_R$ the total number of reactions.

After calculating the direct interaction coefficient for all species pairs, the DRGEP method then performs a graph search---using Dijkstra's algorithm as described by Niemeyer and Sung \cite{Niemeyer:2011fe}---starting at user-selected target species to find the dependency paths for all species from the targets.
A path-dependent interaction coefficient represents the error propagation through a certain path and is defined as the product of intermediate direct interaction coefficients between the target \emph{T} and species \emph{B} through pathway \emph{p}:
\begin{equation}
\label{E:rABp}
	r_{T B, p} = \prod_{j = 1}^{n-1} r_{S_j S_{j + 1}}	,
\end{equation}
where \emph{n} is the number of species between \emph{T} and \emph{B} in pathway \emph{p} and $S_j$ is a placeholder for the intermediate species \emph{j} starting at \emph{T} and ending at \emph{B}.
An overall interaction coefficient is then defined as the maximum of all path-dependent interaction coefficients between the target \emph{T} and each species \emph{B}:
\begin{equation}
\label{E:RAB}
	R_{T B} = \max_{\text{all paths $p$}} \left( r_{T B, p} \right)	.
\end{equation}
Note that in order to solve for the overall interaction coefficients for all species from each target using Dijkstra's algorithm, the traditional shortest-path problem needs to be modified such that the distance of each path is the product rather than sum of intermediate edges, and the ``shortest'' path is that with the maximum value rather than minimum; see Niemeyer and Sung~\cite{Niemeyer:2011fe} for more details.

Pepiot-Desjardins and Pitsch~\cite{Pepiot-Desjardins:2008} also proposed a coefficient scaling procedure to better relate the overall interaction coefficients from different points in the evolution of the reaction system; we adopted this scaling here.
A pseudo-production rate of a chemical element \emph{a} based on the production rates of species containing \emph{a} is defined as
\begin{equation}
\label{E:Patom}
	P_a = \sum_{\text{all species } S} N_{a, S} \max \left( 0, P_S - C_S \right) ,
\end{equation}
where $N_{a, S}$ is the number of atoms \emph{a} in species \emph{S} and $P_S$ and $C_S$ are the production and consumption rates of species \emph{S} as given by Eqs.~\eqref{E:PA} and \eqref{E:CA}, respectively.
The scaling coefficient for element \emph{a} and target species \emph{T} at time \emph{t} is defined as
\begin{equation}
\label{E:alpha_aT}
	\alpha_{a, T} (t) = \frac{ N_{a, T} \left| P_T - C_T \right| }{ P_a } .
\end{equation}
For the set of elements $ \lbrace \mathcal{E} \rbrace $, the global normalized scaling coefficient for target \emph{T} at time \emph{t} is
\begin{equation}
\label{E:alpha_T}
	\alpha_{T} (t) = \max_{a \in \lbrace \mathcal{E} \rbrace} \left( \frac{ \alpha_{a, T} (t) }{ \max\limits_{\text{all time}}  \alpha_{a, T} (t) } \right).
\end{equation}
Given a set of kinetics data $ \lbrace \mathcal{D} \rbrace $ and target species $ \lbrace \mathcal{T} \rbrace $, the overall importance of species \emph{S} to the target species set is
\begin{equation}
\label{E:R_S}
	\overline{ R_S } = \max_{ \substack{ T \in \lbrace \mathcal{T} \rbrace \\ k \in \lbrace \mathcal{D} \rbrace } } \left[ \max_{\text{all time, } k } \left( \alpha_{T} R_{T S} \right) \right].
\end{equation}

The removal of species where $ \overline{ R_S } < \epsilon_{\text{EP}} $ is considered negligible to the overall production\slash consumption rates of the target species and therefore such species are unimportant for the given set of conditions $\lbrace \mathcal{D} \rbrace$ and can be removed from the reaction mechanism.
For a given error limit, an iterative search algorithm chose the optimal threshold $ \epsilon_{\text{EP}} $, taking the following steps.
First, starting with a low value (e.g.,\ 0.001), the algorithm generated an initial skeletal mechanism and calculated the maximum error in ignition delay prediction (compared to the detailed mechanism) for all initial conditions using
\begin{equation}
\label{E:error}
	\delta_{\text{skel}} = \max_{k \in \lbrace \mathcal{D} \rbrace} \frac{ \left | \tau_{\text{det}}^k - \tau_{\text{skel}}^k \right | }{ \tau_{ \text{det} }^k } ,
\end{equation}
where $ \tau_{\text{det}}^k $ and $ \tau_{\text{skel}}^k $ are the ignition delay times calculated by the detailed and skeletal mechanisms for condition \emph{k}, respectively, and $\lbrace \mathcal{D} \rbrace$ is the set of autoignition initial conditions.
If the maximum error for this initial skeletal mechanism is above the user-specified error limit $\delta_{\text{limit}}$, the threshold is decreased.
For this and any subsequent mechanisms, the threshold value increases if the maximum error is below the error limit until the error reaches the specified limit.
This procedure generated a minimal skeletal mechanism using DRGEP for a given error limit prior to application of sensitivity analysis.

\subsection{Sensitivity analysis}
\label{S:SA}

Following the removal of a large number of species by the DRGEP method, a sensitivity analysis algorithm eliminated additional unimportant species.
Species with overall importance values $\left( \overline{ R_S } \right)$ that satisfy $ \epsilon_{EP} < \overline{R_S} < \epsilon^* $, where $ \epsilon^* $ is a higher value (e.g.,\ 0.01--0.2), are classified as ``limbo'' species to be analyzed individually for removal.
Species where $ \overline{R_S} > \epsilon^* $ are classified as ``retained'' species and included in the final skeletal mechanism.

The sensitivity analysis algorithm first evaluates the error induced by the removal of each limbo species, given by
\begin{equation}
\label{E:SA-error}
	\delta_S = \left | \delta_{S, \text{ind}} - \delta_{ \text{skel} } \right | ,
\end{equation}
where $\delta_{\text{skel}}$ is the error of the current skeletal mechanism (prior to temporary removal of limbo species \emph{S}) and $\delta_{S, \text{ind}}$ is the error induced by the removal of limbo species \emph{S} one at a time.
Then, using the criterion given by Eq.~\eqref{E:SA-error}, the algorithm sorts the species for removal in ascending order of induced error.
Species are removed in order until the maximum error reaches the user-specified limit.
By using $\delta_S$ rather than $\delta_{S, \text{ind}}$ directly for the order, species whose removal affect the skeletal mechanism the least are selected first for removal.

\subsection{Unimportant reaction elimination}

After the DRGEPSA method generated a skeletal mechanism, we performed an additional step of further unimportant reaction elimination based on the methodology of Lu and Law~\cite{Lu:2008bi}.
Using an approach based on the CSP importance index~\cite{Lu:2001ve}, the normalized contribution of a reaction \emph{i} to the net production rate of a species \emph{A} is
\begin{equation} \label{E:I_ki}
	I_{A, i} = \frac{ \left| \nu_{A, i} \Omega_i \right| }{ \sum_{j = 1, N_R} \left| \nu_{A, j} \Omega_j \right| } .
\end{equation}
Reactions are considered unimportant if $ I_{A,i} < \epsilon_{\text{reac}} $ for all species \emph{A} in the mechanism; in other words, if
\begin{equation}
	\max_{\text{all species } A} I_{A, i} \ge \epsilon_{\text{reac}} ,
\end{equation}
reaction \emph{i} is retained in the mechanism.
The threshold $\epsilon_{\text{reac}}$ was determined iteratively based on a user-defined error limit in a similar manner to $\epsilon_{\text{EP}}$, described above.

\subsection{Reduction package}

The various reduction stages described above were integrated into the Mechanism Automatic Reduction Software (MARS) package, first described by Niemeyer and coauthors~\cite{Niemeyer:2010bt,Niemeyer:2010}.
MARS performed constant volume autoignition simulations using SENKIN~\cite{Lutz:1997vu} with CHEMKIN-III~\cite{Kee:1996vd} over the range of initial conditions for the desired coverage of the resulting skeletal mechanism.
Using the autoignition results, MARS then densely sampled thermochemical data around the ignition evolution to use in the reduction procedure.
In addition, it used the ignition delay values calculated with the detailed mechanism to measure the error of each skeletal mechanism.
To generate a skeletal mechanism, MARS first applied the DRGEPSA method with a user-defined error limit, the iterative threshold algorithm described in Section~\ref{S:DRGEP}, and a specified $\epsilon^*$ to generate a skeletal mechanism with a minimal number of species.
Next, it performed the unimportant reaction elimination stage to further reduce the complexity of the skeletal mechanism by removing reactions in addition to those removed with unimportant species.

By prescribing the target skeletal mechanism performance through the maximum error limit, rather than a priori selecting the DRGEP cutoff threshold, our approach avoids manual selection of the appropriate threshold, which can vary significantly depending on the detailed mechanism, desired performance, and range of conditions.
The focus on target performance allows many of our results and observations, particularly regarding the interaction between fuel components to be discussed below, to remain general.

\section{Results and discussion}
\label{S:results}

In the following sections, we describe and discuss our results.
First, we generated skeletal mechanisms for a TRF mixture at different error limits and ranges of initial conditions, and compared the performance of these for autoignition.
We then extended the validation of these skeletal mechanisms to phenomena other than that used in the reduction; specifically, we considered perfectly stirred reactor (PSR) and laminar flame propagation simulations.
Next, we compared different reduction approaches for generating skeletal mechanisms for multicomponent fuels: combining separately produced skeletal mechanisms for the neat fuel components, and reducing the multicomponent mixture together.
In both approaches, we used the same error limit for a fair comparison.
After this, we studied the performance of the skeletal mechanisms in autoignition with varying mixture composition.
Finally, in addition to validating using global properties, we added an additional level of rigor and analyzed the prediction of species mass fraction profiles in HCCI simulations.

\subsection{Skeletal mechanisms and validation}

We generated skeletal mechanisms at varying levels of detail from the detailed mechanism for TRF mixtures (n-heptane, iso-octane, and toluene) of Mehl et al.~\cite{Mehl:2011cn}, containing \num{1389} species and \num{5935} reactions, using DRGEPSA followed by further unimportant reaction elimination.
Two sets of constant volume autoignition initial conditions, both covering 1--\SI{20}{\atm} and equivalence ratios 0.5--1.5, were used to generate chemical kinetics data: 1) 700--\SI{1600}{\kelvin} and 2) \num{1000}--\SI{1600}{\kelvin}; the former is referred to as the comprehensive range and the latter the high-temperature range.
These conditions were chosen to match those typically used for the development of detailed and reduced mechanisms~\cite{Chaos:2007fc,Blanquart:2009iw,Niemeyer:2010bt,Luo:2010cu,Hansen:2011bu,Blurock:2013}.
The comprehensive range of conditions includes complex low-temperature chemistry and offers a broad range applicability at the cost of a typically larger mechanism size, while a high-temperature mechanism suitable for, e.g., flame studies neglects these effects and as such can contain smaller numbers of species and reactions.
Caution should be exercised in applying either of the resulting skeletal mechanisms outside their respective thermal condition ranges.

Early in our efforts we discovered an inconsistency in PSR calculations using the detailed mechanism of Mehl et al.~\cite{Mehl:2011cn}, where at $\phi = 1.0$ it predicted higher temperatures at large residence times approaching equilibrium than the adiabatic flame temperature determined via equilibrium calculations.
Further analysis revealed the cause of this inconsistency to be a dead-end pathway in the toluene submechanism: specifically, a fuel breakdown pathway dead-ending at \ce{nC4H3}, a radical that may be important for aromatic formation pathways~\cite{Miller:1990et,Richter:2000uy,Hansen:2006hj,Lories:2010gp}.
In the current TRF mechanism, this species is produced through the dissociation of benzyl radical, and only participates in one other reaction---combining with propargyl (\ce{C3H3}) to form fulvenallene (\ce{C7H6}).
Removing \ce{nC4H3} and the two reactions in which it participates corrected the aforementioned discrepancy in equilibrium temperatures for $\phi = 1.0$, and we used the resulting modified mechanism in all the analysis and calculations that follow.
However, this issue warrants further development of the detailed mechanism, in particular the aromatic chemistry.

The mixture composition we used consists of 60.54\slash 20.64\slash 18.82\% (by liquid volume) toluene\slash iso-octane\slash n-heptane, or 69.06\slash 15.38\slash 15.56\% by molar percentage, taken from Morgan et al.~\cite{Morgan:2010}, which corresponds to gasoline with a research octane number (RON) of 95 and motor octane number (MON) of 85.
We refer to this mixture as TRF1.
We selected error limits of 10\% and 30\%, and used n-heptane, iso-octane, toluene, oxygen, and the inert nitrogen (to prevent removal) for the DRGEP target species.
As in previous efforts using DRGEPSA~\cite{Niemeyer:2010bt}, we used $\epsilon^* = 0.1$ to determine the upper limit of limbo species considered by the sensitivity analysis portion of DRGEPSA.

The skeletal mechanism reduction results are summarized as follows.
For the comprehensive range of conditions, the DRGEP algorithm selected $\epsilon_{\text{EP}}$ values of \num{5e-3} and \num{1e-2} for the 10\% and 30\% error limits, removing 902 and 988 species, respectively.
The sensitivity analysis stages then removed a further 101 and 125 species each.
The unimportant reaction elimination stage, using $\epsilon_{\text{reac}}$ thresholds of \num{6e-3} and \num{3e-2}, removed 347 and 386 reactions, respectively.
The final 10\%- and 30\%-error comprehensive skeletal mechanisms consisted of 386 species and \num{1591} reactions, and 276 species and 936 reactions, respectively.
For the high-temperature conditions, DRGEP thresholds $\epsilon_{\text{EP}}$ of \num{8e-3} and \num{1.6e-2} removed \num{1103} and \num{1160} species for the 10\%- and 30\%-error limits, respectively, while the sensitivity analysis stages eliminated 87 and 56 species.
The unimportant reaction elimination stages selected cutoff thresholds $\epsilon_{\text{reac}}$ of \num{1e-3} and \num{2e-2} to remove 155 and 334 reactions for the 10\%- and 30\%-error limits, respectively, resulting in final skeletal mechanisms consisting of 199 species and \num{1011} reactions, and 173 species and 689 reactions, respectively.

Figures \ref{F:comp-ignition-delay} and \ref{F:high-ignition-delay} show the comparisons of the predicted ignition delay times for the comprehensive and high-temperature skeletal mechanisms, respectively, with those from the detailed mechanism.
Note that each set of skeletal mechanisms was validated over their respective target range of conditions, where the associated error limits used in the reduction procedure constrained the error in ignition delay.
All four skeletal mechanisms accurately predicted the ignition delays of the detailed mechanism, within the specified error limits.

Next, we performed extended validation of the skeletal mechanisms in phenomena other than that used for the reduction procedure: PSR and one-dimensional laminar flame propagation simulations.
Figure \ref{F:comp-psr} shows the PSR temperature response curves as a function of residence time of the detailed mechanism and both comprehensive skeletal mechanisms for pressures of 1 and \SI{20}{\atm}, equivalence ratios of 0.5--1.5, and an inlet temperature of \SI{300}{\kelvin}.
We used the same TRF1 mixture composition as in the mechanism reduction.
Only the skeletal mechanism generated using a 10\% error limit closely reproduced the temperature response curves and extinction turning points of the detailed mechanism over the full range of conditions considered.
The smaller 30\%-error skeletal mechanism performed adequately for $\phi = 1.5$, showing some error in the turning points, as well as the $\phi = 1.0$ and \SI{1}{\atm} case, but demonstrated noticeable errors in the remaining conditions.
In particular, this mechanism poorly reproduced the temperature response curve for $\phi = 0.5$, and severely underpredicted the temperature along the upper branch for the $\phi = 1.0$ and \SI{20}{\atm} case (although it closely matched the turning point).

In order to determine why the 30\%-error skeletal mechanism performed poorly in PSR simulations when a skeletal mechanism for neat n-decane generated using the same procedure with the same parameters---30\% error limit and $\epsilon^* = 0.1$---performed well in both PSR and one-dimensional laminar flame simulations in our previous work~\cite{Niemeyer:2010bt}, we generated skeletal mechanisms using these reduction parameters for each of the three components: n-heptane, iso-octane, and toluene.
These reductions used the same comprehensive range of conditions as described previously.
The resulting skeletal mechanisms for n-heptane, iso-octane, and toluene consisted of 177 species and 730 reactions, 198 species and 769 reactions, and 60 species and 185 reactions, respectively.
Based on the 30\% error limit, our reduction algorithm selected DRGEP and unimportant reaction cutoff thresholds ($\epsilon_{\text{EP}}$ and $\epsilon_{\text{reac}}$) of \num{1.2e-2} and \num{1e-2}, \num{4e-3} and {\num{2e-2}, and \num{7e-3} and 0.13 for n-heptane, iso-octane, and toluene, respectively.

Figures \ref{F:psr-nheptane}, \ref{F:psr-isooctane}, and \ref{F:psr-toluene} show the PSR validation of the n-heptane, iso-octane, and toluene skeletal mechanisms, respectively.
All three mechanisms closely reproduced the ignition delays for each neat fuel calculated by the detailed mechanism, bounded by the 30\% error limit, so we omit these results here.
The n-heptane and iso-octane mechanisms closely reproduced the temperature response curves and extinction turning points calculated by the detailed mechanisms, with small errors near the turning points for some conditions.
However, the 60-species toluene skeletal mechanism performed more poorly, especially for $\phi = 0.5$.
Although not shown here, the skeletal mechanism resulting from the DRGEP stage closely reproduces the PSR temperature response curves; therefore, species important to PSR were removed during the sensitivity analysis stage, resulting in the observed errors.
We therefore generated a second toluene skeletal mechanism using a smaller $\epsilon^*$ value of 0.05 to further limit species considered by sensitivity analysis, resulting in a slightly larger mechanism with 69 species and 274 reactions.
The performance of this skeletal mechanism is also shown in Fig.~\ref{F:psr-toluene}; compared to the initial toluene skeletal mechanism, this mechanism showed improved performance in calculating the turning points for most conditions and for $\phi = 0.5$ in particular.

The errors demonstrated by the skeletal mechanism for neat toluene generated using $\epsilon^* = 0.1$ suggested that the presence of toluene caused some of the large errors in the 30\%-error comprehensive skeletal mechanism for the TRF blend.
This was likely due to the presence of different pathways initiated during ignition and the steady burning present in PSR conditions.
Using error in ignition delay alone may not be able to capture the importance of some species in other phenomena such as PSR, therefore requiring limitations on the limbo species.
Furthermore, accurately reproducing ignition delay does not guarantee the same level of performance in PSR when using a sensitivity analysis stage.
However, the greater extent of errors seen in Fig.~\ref{F:comp-psr} compared to those in Fig.~\ref{F:psr-toluene} indicated that moving from a single fuel to a multicomponent mixture while retaining the same 30\% error limit might also require a smaller $\epsilon^*$ value.
Motivated by this result, we used the same smaller $\epsilon^*$ value of 0.05 combined with the 30\% error limit and generated another comprehensive TRF skeletal mechanism.
Since we only changed $\epsilon^*$, the DRGEP stage behaved the same, but the sensitivity analysis stage removed only 100 species in this case, compared to 125 when $\epsilon^* = 0.1$.
The unimportant reaction elimination stage used $\epsilon_{\text{reac}} = \num{2e-2}$ to remove 345 reactions, resulting in a final skeletal mechanism with 301 species and \num{1192} reactions---slightly larger than the original 30\%-error limit mechanism with 276 species and 936 reactions.
We omit the performance of this skeletal mechanism in autoignition because the results were nearly identical to those already shown in Fig.~\ref{F:comp-ignition-delay}.
Figure~\ref{F:comp-psr-05} shows the performance in PSR; this mechanism closely reproduced the temperature response curves for all conditions.

These results suggest that when using error in ignition delay to perform the sensitivity analysis, a larger $\epsilon^*$ value can be paired with a larger error limit for normal and branched alkanes like n-heptane and iso-octane, as well as n-decane as demonstrated previously by Niemeyer et al.~\cite{Niemeyer:2010bt}.
However, mixtures with toluene---and multicomponent fuel mixtures in general---may require either a tight error limit (e.g., $\leq 10\%$) or a smaller $\epsilon^*$ value for good performance in PSR simulations.
Unfortunately, both strategies will result in larger skeletal mechanisms.

Another solution may be to sample PSR data in addition to autoignition and use the extinction residence time as another measure of error, as in the efforts of Lu and Law~\cite{Lu:2005ce,Lu:2006bb,Lu:2008bi} and Luo and coworkers~\cite{Luo:2010cu,Luo:2012cr,Luo:2012dp}.
However, many skeletal reduction approaches use autoignition simulations alone as the basis for the reduction procedure, e.g.,~\cite{Valorani:2006bp,Valorani:2007ef,Pepiot-Desjardins:2008,Mitsos:2008,Prager:2009ge,Niemeyer:2010bt,Shi:2010a,Zsely:2011im,Mehl:2011jn,Bahlouli:2012kq}.
Whether including PSR simulations in the reduction procedure---combined with using a larger $\epsilon^*$ value (e.g., 0.5--0.9)---will result in a more compact but equally well-performing skeletal mechanism compared to using autoignition alone with a small $\epsilon^*$ requires further investigation and will be explored in future work.

Figure \ref{F:high-psr} shows the comparisons of the PSR results between the high-temperature skeletal mechanisms and detailed mechanism.
In this case, both skeletal mechanisms accurately reproduced the temperature response curves and turning points of the detailed mechanism for the conditions considered.
The only noticeable error in extinction turning point occurred in the $\phi = 1.5$, \SI{20}{\atm} case for the 30\%-error mechanism: approximately 21\%, under the set error limit for autoignition.

Figures \ref{F:comp-flame-speed} and \ref{F:high-flame-speed} show the laminar flame speeds calculated using the detailed mechanism compared against those from the comprehensive and high-temperature skeletal mechanisms, respectively, which were computed using the PREMIX program in the CHEMKIN-PRO package~\cite{chemkin:2012}.
Here, we only used the larger of the two 30\%-error comprehensive skeletal mechanisms, generated using $\epsilon^* = 0.05$---as in the case of the PSR simulations, the corresponding smaller skeletal mechanism created using $\epsilon^* = 0.1$ performed poorly here.
For these calculations, we considered the effects of thermal diffusion (the Soret effect), and used the mixture-averaged diffusion model.
Further, we enabled the velocity correction option to ensure mass conservation.
All four skeletal mechanisms accurately predicted the laminar flame speeds at both 1 and \SI{20}{\atm}.
At worst, the 10\%- and 30\%-error comprehensive skeletal mechanism overpredicted the laminar flame speeds of the detailed mechanism by 3.8\% and underpredicted by 5.5\%, respectively; the largest errors of the high-temperature skeletal mechanisms were 5.4\% and 11.2\% for the 10\%-error and 30\%-error mechanisms, respectively.

\subsection{Comparison of multicomponent reduction approaches}

Using the various skeletal mechanisms described in the previous section, we next compared the efficacy of two approaches to generating skeletal mechanisms for multicomponent fuel blends: (1) reducing the multicomponent fuel mixture together; and (2) first generating skeletal mechanisms for the neat components separately, then combining these together to create the reduced multicomponent mechanism.
We refer to these as the ``multicomponent'' and ``combination'' reduction approaches, respectively, in the rest of this section.
The multicomponent TRF skeletal mechanism corresponding to a 30\%-error limit and $\epsilon^* = 0.05$ contained 301 species and \num{1192} reactions.
Merging the 30\%-error-limit skeletal mechanisms for neat n-heptane (177 species and 730 reactions), iso-octane (198 species and 769 reactions), and toluene (69 species and 274 reactions) resulted in a combined TRF skeletal mechanism with 355 species and \num{1416} reactions.
Note that the n-heptane and iso-octane skeletal mechanisms were generated using $\epsilon^* = 0.1$, while the toluene mechanism used $\epsilon^* = 0.05$.

Immediately, we observed that the combination approach produced a larger skeletal mechanism, with an additional 54 species and 224 reactions.
However, this larger size may be acceptable depending on the mechanism performance, so we next calculated ignition delays of TRF mixtures for various initial temperatures, pressures, and equivalence ratios using the combined TRF skeletal mechanism.
Figure~\ref{F:combined-TRF-igndelay} shows the comparison of ignition delay calculations for the two skeletal mechanisms and the detailed mechanism.
The multicomponent skeletal mechanism with 301 species predicted ignition delay over the full range of conditions within the 30\% error limit (with a maximum error of 28.9\%), but the combination TRF skeletal mechanism violated this limit with a maximum error of 100.7\%.
Note that the largest errors of both skeletal mechanisms occurred at lower temperatures, but for the multicomponent skeletal mechanism this error remained bound by the 30\% reduction limit.
The (larger) combination skeletal mechanism, on the other hand, overpredicted ignition delay substantially at the lowest temperatures and high pressure, particularly for stoichiometric-to-rich equivalence ratios.
In addition, noticeable errors are observed around \SI{850}{\kelvin} for atmospheric pressure.

In order to determine the reason for the poor performance of the combination skeletal mechanism compared to that of the multicomponent skeletal mechanism, we studied the species and reactions present in the latter but missing from the former.
In fact, even though the combination mechanism contained 54 additional species and 224 additional reactions, it was missing 40 species and 222 reactions present in the smaller multicomponent skeletal mechanism---including some important cross reactions between the three component fuels.
Specifically, eight \ce{H} abstraction cross-reactions between toluene and n-heptane\slash iso-octane were not included in the combination mechanism:
\begin{equation*}
\ce{RH + Q <-> R + QH} \tag{R3} \label{reac:Habs},
\end{equation*}
where \ce{RH} refers to toluene and \ce{QH} refers to n-heptane or iso-octane; Sakai et al.~\cite{Sakai:2009bp} indicated these cross reactions can be important in mixtures of toluene and n-heptane\slash iso-octane.
In addition, two phenyl radical (\ce{C6H5}) to alkene addition reactions were missing, which Sakai et al.~\cite{Sakai:2009bp} indicated can also have large effects in mixtures of toluene and iso-octane.

In order to test this hypothesis and better understand the interactions among the three fuel components, we added back the reactions described above.
Returning the two addition reactions of phenyl radical to alkene did not affect the performance of the skeletal mechanism, but returning the eight \ce{H}-abstraction reactions reduced the maximum error to 30.5\%, or nearly the performance of the multicomponent skeletal mechanism.
Our results indicate the high importance of these cross reactions for predicting ignition delay in mixtures of toluene and the PRFs, particularly at low temperatures and high pressures.

These results suggest that generating a skeletal mechanism for multicomponent fuel mixtures by combining skeletal mechanisms for the neat components may result in (1) a larger mechanism compared to performing the reduction for the blend directly and (2) the loss of important interactions between the fuel components.
The latter trend will likely be exacerbated for multicomponent blends with even greater kinetic interactions between the fuel components than the TRF used here.
Therefore, we recommend performing the mechanism reduction for the multicomponent blend rather than combining skeletal mechanisms for neat fuels as in previous studies~\cite{Andrae:2008,Blanquart:2009iw,Ra:2011gh}.

In a final note, while the multicomponent skeletal mechanism performed acceptably for TRF autoignition, PSR, and laminar flame calculations, performance for the neat components in the same phenomena is not guaranteed since only multicomponent conditions were used for the reduction.
On the other hand, the mechanism formed by combining the neat-fuel skeletal mechanisms does retain this capability---though it could not accurately capture the multicomponent blend behavior, as shown.
If the ability to predict targets for the individual components is desired in the multicomponent skeletal mechanism, then these targets should be included in the reduction process.
However, this may result in a larger skeletal mechanism.
We performed an additional reduction using the above multistage procedure, with the comprehensive initial condition range for the multicomponent TRF1 mixture as well as neat n-heptane, iso-octane, and toluene.
To compare with the skeletal mechanisms discussed in this section, we set the error limit at 30\% and $\epsilon^* = 0.1$.
The mechanism reduction algorithms selected $\epsilon_{\text{EP}} = \num{7e-3}$ and $\epsilon_{\text{reac}} = \num{1e-2}$, resulting in a skeletal mechanism with 576 species and 2658 reactions.
Validation of this mechanism showed excellent agreement with the detailed mechanism for autoignition, PSR, and laminar flame calculations (see the full validation results in the supplementary material); however, this fidelity came at the cost of a large number of species and reactions.
For the same error limit, this mechanism contains more than 200 additional species compared to the combination approach and 300 more species than the multicomponent reduction approach.

In the following section, we describe the performance of the multicomponent skeletal mechanisms---generated at one fuel blend composition---when varying the fuel mixture in autoignition simulations as a demonstration; this includes conditions for the neat fuel components.

\subsection{Performance with varying mixture composition}

Next, we studied the performance of the skeletal mechanisms in constant volume autoignition simulations with varying mixture composition.
For the comprehensive skeletal mechanisms---which contain low-temperature chemistry---we used initial conditions representative of RON and MON tests, taken from Morgan et al.~\cite{Morgan:2010}.
The RON-like case initiated at \SI{800}{\kelvin} and \SI{23}{\bar} and the MON-like case at \SI{900}{\kelvin} and \SI{20}{\bar}.
Both cases used a stoichiometric mixture, based on the particular composition of the ternary blend.

Figures \ref{F:comp10-MON} and \ref{F:comp10-RON} show the error in ignition delay using the 10\%-error skeletal mechanism for the MON- and RON-like initial conditions as a function of the mixture composition varying over the three-component space.
In both cases, the mechanism performed best closest to the TRF1 mixture composition used for the reduction, and in general the error remained around 10\% over most of the mixture space, although the error increased past the 10\% limit as the mixture composition moved towards neat iso-octane and n-heptane.
Figures \ref{F:comp30-MON} and \ref{F:comp30-RON} show the ignition delay error for the 30\%-error skeletal mechanism for the MON- and RON-like cases, respectively.
For the MON-like case, as the mixture compositions moved away from the TRF1 point, the error increased substantially beyond the 30\% limit.
The error in the RON-like ignition delay showed less extreme increases, although it became high for mixtures with small amounts of toluene and large amounts of iso-octane.

For both skeletal mechanisms in this case, we attributed the increase in error (to varying degrees) to the removal of species and reactions important to iso-octane, the smallest fuel component in the TRF1 mixture.
In the case of the larger 10\%-error mechanism, the error only became unacceptably large for mixtures comprised of mostly iso-octane; a tight error limit appears to ensure wide-ranging applicability.
On the other hand, the smaller 30\%-error skeletal mechanism performed well only near the mixture composition used in the reduction, suggesting that a greater extent of reduction comes at the cost of wide-ranging applicability.

The high-temperature skeletal mechanisms lacked the low-temperature chemistry necessary to accurately predict ignition delay in the MON- and RON-like cases, so we instead used initial conditions of \SI{1200}{\kelvin} and \SI{10}{\atm}, again with a stoichiometric mixture.
Figure \ref{F:high10} shows the error in ignition delay with a varying fuel composition using the 10\%-error skeletal mechanism with 199 species.
In this case, while the error increased slightly moving away from the TRF1 mixture, it remained acceptable even for neat iso-octane.
The 30\%-error skeletal mechanism behaved similarly to the 10\%-error mechanism, with results similar to those shown in Fig.~\ref{F:high10}.

For high temperatures, skeletal mechanisms depend less on the fuel composition used for the reduction procedure, due to the commonality of species and reactions important to high-temperature oxidation (e.g., \ce{H + O2} branching for the current temperature\slash pressure combination).
However, the low-temperature chemistry involves more fuel-specific competing pathways (e.g., \ce{R + O2} pathways~\cite{Battin-Leclerc:2008,Zador:2011kz}), so the species retained by the skeletal reduction depend more on the particular composition used.
Based on the results shown here, skeletal mechanisms for multicomponent fuels that consider low-temperature chemistry should be applied only near the composition used for the reduction---although a tight error limit may ensure acceptable performance over a wide range of mixtures at the cost of a larger mechanism.

\subsection{Detailed validation of species profiles}

Matching global properties such as ignition delay and extinction residence time does not guarantee agreement of time-varying or spatially resolved scalars such as species mass fractions.
Therefore, we performed detailed validation of the two comprehensive skeletal mechanisms with HCCI engine simulations, using the CHEMKIN-PRO internal combustion engine simulator~\cite{chemkin:2012}.
The simulated engine properties, taken from Sj\"{o}berg et al.~\cite{Sjoberg:2007}, included a displacement volume of \SI{0.981}{\liter}, a compression ratio of 14, and a connecting rod to crank radius ratio of 3.2.
We selected five sets of initial conditions, given in Table~\ref{T:engine-IC}, modified from those used by Andrae et al.~\cite{Andrae:2008}.

Both comprehensive skeletal mechanisms performed well for all the initial conditions considered, with the 10\%- and 30\%-error skeletal mechanisms predicting the ignition delay for all five cases with errors of less than approximately 2.1\% and 8.2\%, respectively.
Furthermore, both mechanisms closely matched the mass fraction profiles of the major species, as shown for case IC5 in Fig.~\ref{F:IC5-10} for the 10\%- and in Fig.~\ref{F:IC5-30} for 30\%-error skeletal mechanisms.
Figures~\ref{F:IC5-10} and \ref{F:IC5-30} show species mass fractions in terms of both original crank angle degrees (\textdegree CA) after top dead center (ATDC) and shifted \textdegree CA based on ignition-delay timing; the close agreement is more evident in the latter.
For this case, the 10\%- and 30\%-error skeletal mechanisms predicted ignition within 1.1 and 3.2 \textdegree CA, respectively.
Both mechanisms performed similarly well in the other four cases, so we omit the corresponding plots here for brevity---they can be found in the supplementary material.

\section{Conclusions}
\label{S:conclusions}

In the current work, we developed and studied skeletal mechanisms for a three-component toluene reference fuel, and in the process of doing so discussed some issues with, and corresponding strategies for, mechanism reduction of multicomponent fuels in general.
Our reduction approach consisted of three stages: (1) directed relation graph with error propagation (DRGEP), followed by (2) sensitivity analysis (SA), and finally (3) unimportant reaction elimination.
Using this multistage reduction approach, we generated skeletal mechanisms corresponding to error limits of 10\% and 30\% valid over a comprehensive range of conditions consisting of 386 species and \num{1591} reactions, and 301 species and 1192 reactions, respectively.
In addition, by limiting the range of conditions to high-temperatures, we constructed more compact skeletal mechanisms with 199 species and \num{1011} reactions, and 173 species and 689 reactions, corresponding to 10\%- and 30\%-error limits, respectively.
The skeletal mechanisms were validated using autoignition simulations, and all four well predicted the ignition delays of the detailed mechanism for their corresponding ranges of conditions.

We also performed extended validation by studying the performance of the skeletal mechanisms in (1) perfectly stirred reactor (PSR) and laminar flame simulations; (2) autoignition simulations with a varying mixture composition; and (3) single-zone homogeneous charge compression ignition engine simulations, where the prediction of species mass fractions were considered.
Furthermore, we compared the efficacy of generating a direct multicomponent skeletal mechanism to combining skeletal mechanisms for neat fuel components.
In doing so, we drew the following conclusions:
\begin{itemize}
	\item Either a tight error limit or a smaller cutoff value for the species considered by sensitivity analysis was needed to ensure species important to phenomena outside autoignition, such as PSR and laminar flame propagation, were retained.
	\item Compared to performing the reduction for the multicomponent mixture, combining skeletal mechanisms generated for the neat fuels (and using the same error tolerances) resulted in a larger skeletal mechanism that lacked important cross reaction pathways between the fuel components, noticeably increasing error in ignition delay prediction for fuel blends at lower temperatures and higher pressures.
	\item Performing a reduction using both multicomponent and neat fuel component conditions ensures high-fidelity with the detailed mechanism for both, but results in a substantially larger skeletal mechanism.
	\item In general, skeletal mechanisms for multicomponent fuels should be applied near the mixture composition used to generate them. Otherwise, significant errors can be encountered as the mixture composition moves away from that point.
	\item Alternatively, tight error control can be used to ensure widely applicable skeletal mechanism, at the cost of a larger size.
	\item Accurately predicting global properties such as ignition delay does not guarantee predictability of time-varying  scalars such as species profiles. Error limits need to be judiciously chosen based on the problem and phenomena of interest.
\end{itemize}

Finally, we note the large size of the mechanisms developed here, even those generated using a less-stringent error limit.
Unfortunately, the strategies for dealing with the issues we discussed---using a tight error limit, or limiting the limbo species considered for sensitivity analysis---in general will result in larger mechanisms that, while representing notable reduction from the starting detailed mechanism, still pose significant computational challenges for use in multidimensional combustion simulations.
Additional reduction stages can be applied to further reduce the size, such as isomer lumping and the quasi-steady-state approximation~\cite{Lu:2008bi,Lu:2009gh}, but these stages typically only refine the size of the mechanism rather than remove a significant number of species.

In general, dealing with the issues we encountered required a tighter error limit or otherwise preventing species from being removed, resulting in larger mechanism sizes.
With this and the aforementioned limitations of additional reduction stages in mind, rather than focusing on a priori mechanism reduction, adaptive\slash dynamic reduction approaches (e.g., \cite{Liang:2009,Liang:2009a,Shi:2010,Gou:2013eu,Yang:2013ip}) may offer a more practical option for handling large detailed mechanisms for multicomponent fuels.
These approaches use the local thermochemical state during a simulation to produce a locally relevant skeletal mechanism---avoiding many of the issues shown in the current work.
However, additional research on these methods is still needed, particularly on error control~\cite{Gou:2013eu}.


\section*{Acknowledgments}

This work was supported by the National Science Foundation under Grant No.\ 0932559 and through the National Science Foundation Graduate Research Fellowship under Grant No.\ DGE-0951783, the Department of Defense through the National Defense Science and Engineering Graduate Fellowship program, and the Combustion Energy Frontier Research Center---an Energy Frontier Research Center funded by the US Department of Energy, Office of Science, Office of Basic Energy Sciences under award number DE-SC0001198.
The authors would also like to thank Drs.\ William Pitz and Marco Mehl of LLNL for providing the TRF detailed mechanism, and Prof.\ Tianfeng Lu for providing his PSR program. We also gratefully thank the reviewers for their helpful comments on the first version of this article.

\section*{Supplementary material}
All multicomponent skeletal mechanisms generated in this work are available as supplemental material, as well as the omitted validation figures.
The MARS reduction package is available upon request to the authors.



\pagebreak

\begin{table}[htb]
\begin{center}
\begin{tabular}{c c c c c}
\toprule
& 	$\phi$	& speed (rpm)	& $T_0$ (K)	& $p_0$ (bar) \\
\midrule
IC1	&	0.75	& 900	&	447	&	1.7 \\
IC2	&	0.5		& 900	& 	421	&	3.2 \\
IC3	&	0.75	& 900	& 	489	&	1.6 \\
IC4	&	0.5		& \num{1200}	&	489	&	1.5 \\
IC5	&	0.5		& \num{1200}	&	435	&	3.3 \\
\bottomrule
\end{tabular}
\caption{Initial conditions used for compression-ignition engine simulations, modified from those given by Andrae et al.~\cite{Andrae:2008}. Simulations initiate at 99~\textdegree CA before top dead center (BTDC) at temperature $T_0$ and pressure $p_0$.}
\label{T:engine-IC}
\end{center}
\end{table}


\clearpage


\listoffigures

\begin{figure}[htbp]
\centering
\includegraphics[width=0.65\textwidth]{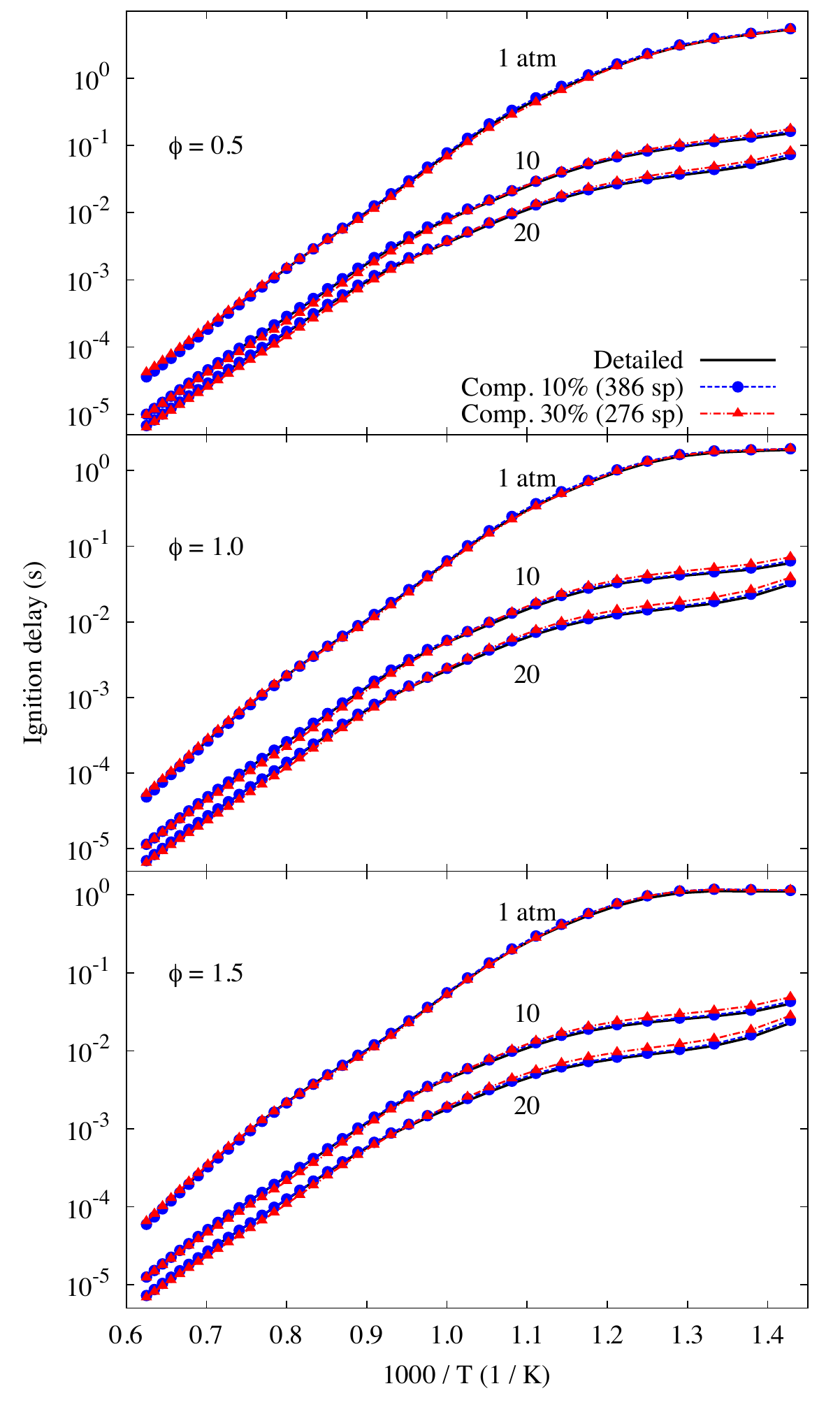}
\caption{Comparisons of the predicted ignition delay with the detailed TRF mechanism and the two comprehensive skeletal mechanisms for initial conditions covering 700--\SI{1600}{\kelvin}, 1--\SI{20}{\atm}, and equivalence ratios of 0.5--1.5.}
\label{F:comp-ignition-delay}
\end{figure}

\begin{figure}[htbp]
\centering
\includegraphics[width=0.65\textwidth]{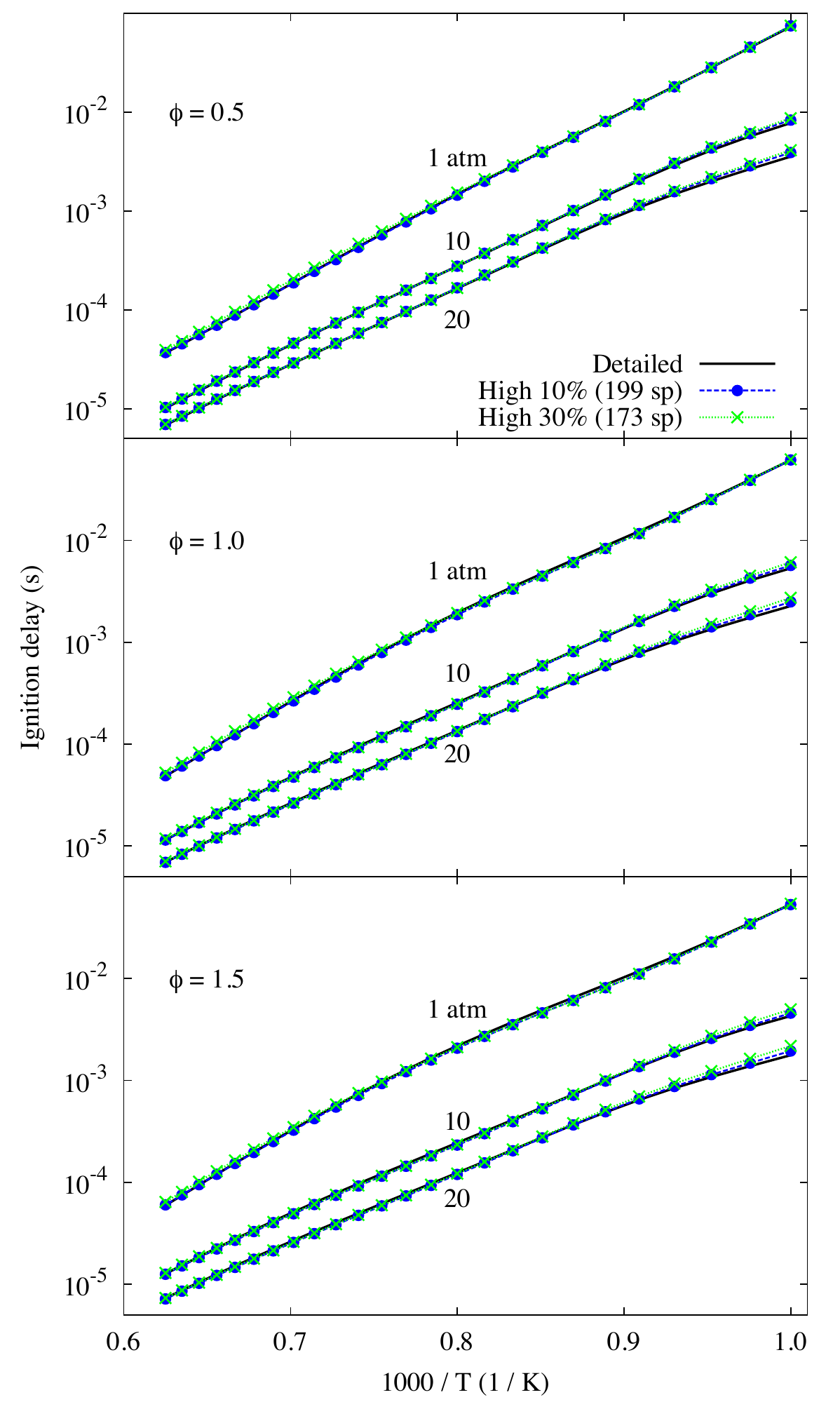}
\caption{Comparisons of the predicted ignition delay with the detailed TRF mechanism and the two high-temperature skeletal mechanisms for initial conditions covering \num{1000}--\SI{1600}{\kelvin}, 1--\SI{20}{\atm}, and equivalence ratios of 0.5--1.5.}
\label{F:high-ignition-delay}
\end{figure}

\begin{figure}[htbp]
\centering
\includegraphics[width=\textwidth]{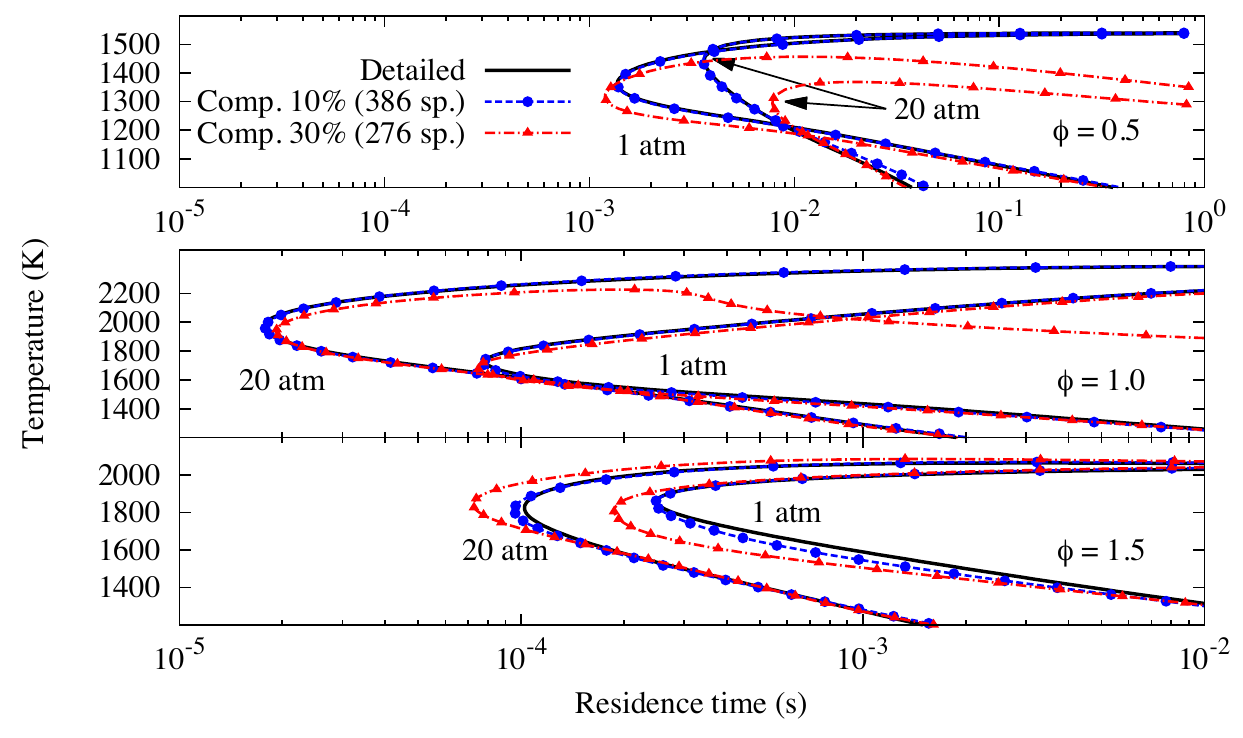}
\caption{Comparisons of PSR temperature response curves for the detailed TRF mechanism and the two comprehensive skeletal mechanisms for pressures of 1 and \SI{20}{\atm}, equivalence ratios of 0.5--1.5, and an inlet temperature of \SI{300}{\kelvin}.}
\label{F:comp-psr}
\end{figure}

\begin{figure}[htbp]
\centering
\includegraphics[width=\textwidth]{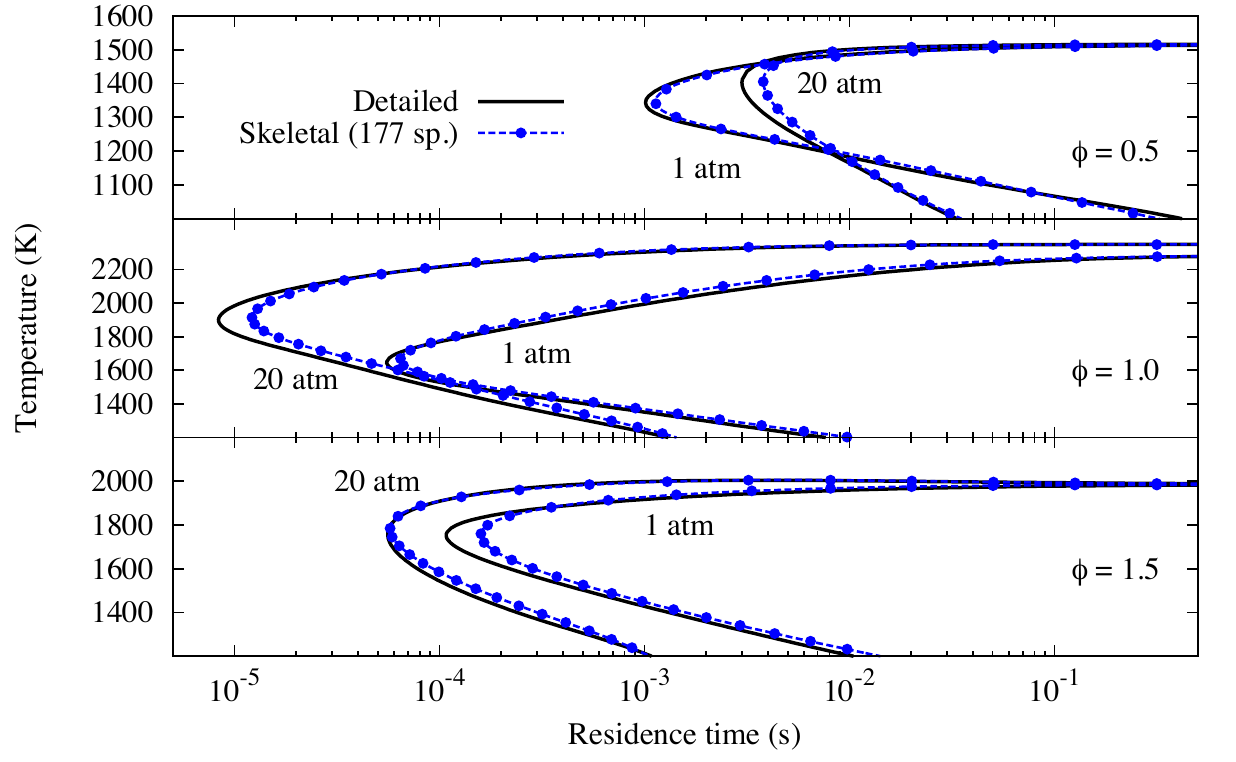}
\caption{Comparisons of PSR temperature response curves for the detailed mechanism and n-heptane skeletal mechanism (30\% error limit, 177 species and 730 reactions) for pressures of 1 and \SI{20}{\atm}, equivalence ratios of 0.5--1.5, and an inlet temperature of \SI{300}{\kelvin}.}
\label{F:psr-nheptane}
\end{figure}

\begin{figure}[htbp]
\centering
\includegraphics[width=\textwidth]{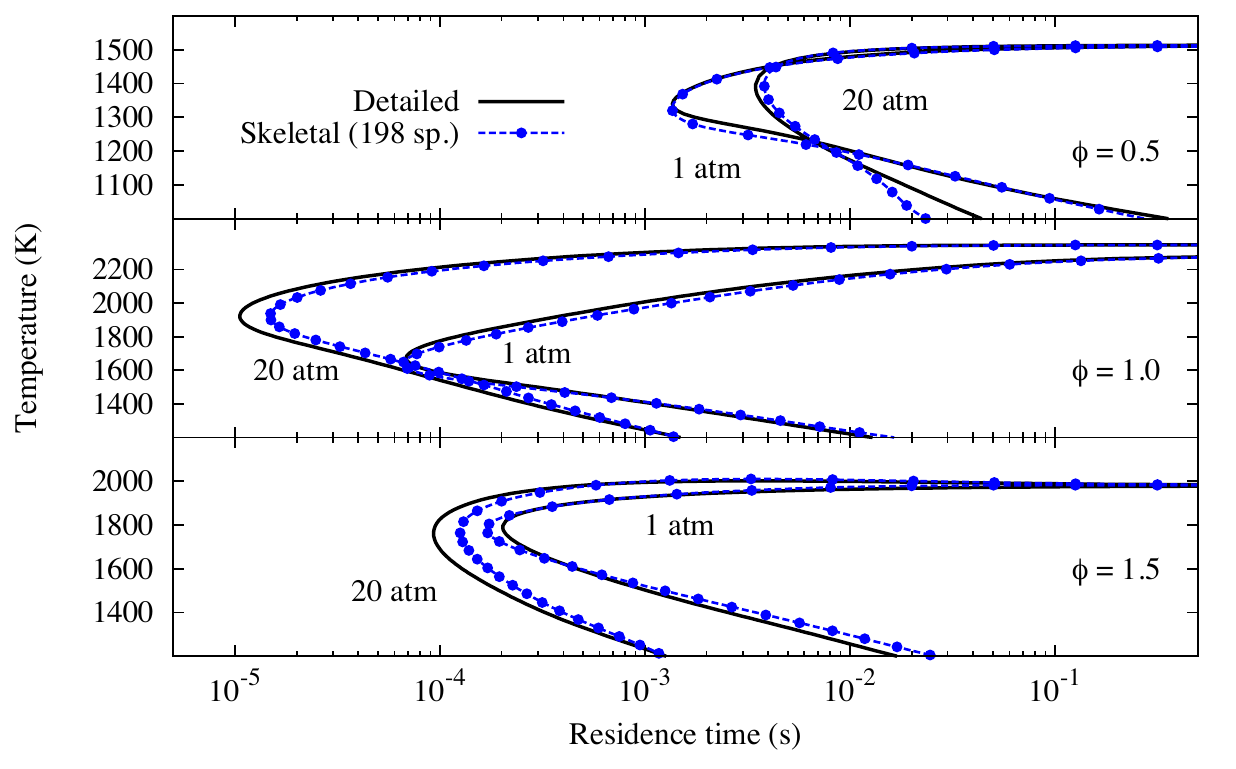}
\caption{Comparisons of PSR temperature response curves for the detailed mechanism and iso-octane skeletal mechanism (30\% error limit, 198 species and 769 reactions) for pressures of 1 and \SI{20}{\atm}, equivalence ratios of 0.5--1.5, and an inlet temperature of \SI{300}{\kelvin}.}
\label{F:psr-isooctane}
\end{figure}

\begin{figure}[htbp]
\centering
\includegraphics[width=\textwidth]{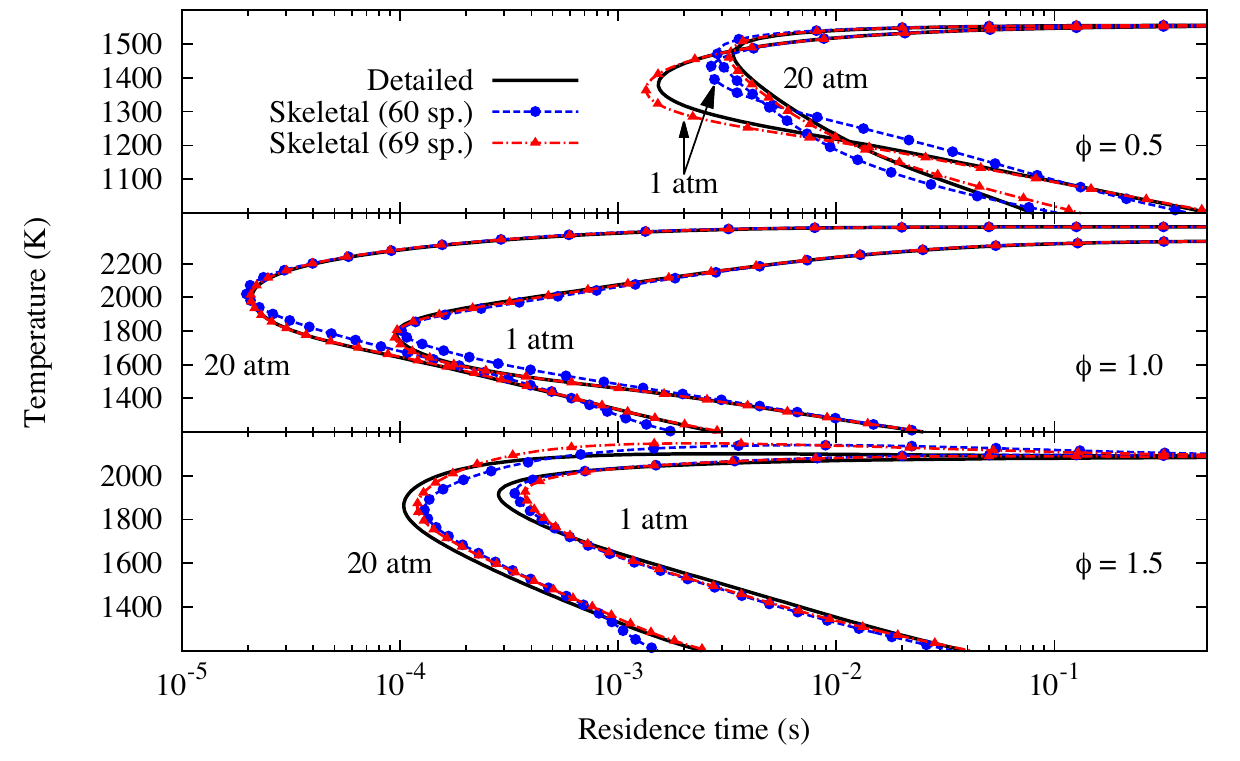}
\caption{Comparisons of PSR temperature response curves for the detailed mechanism and toluene skeletal mechanisms generated using a 30\% error limit and $\epsilon^*$ values of 0.1 and 0.05, corresponding to 60 species and 185 reactions, and 69 species and 274 reactions, respectively, for pressures of 1 and \SI{20}{\atm}, equivalence ratios of 0.5--1.5, and an inlet temperature of \SI{300}{\kelvin}.}
\label{F:psr-toluene}
\end{figure}

\begin{figure}[htbp]
\centering
\includegraphics[width=\textwidth]{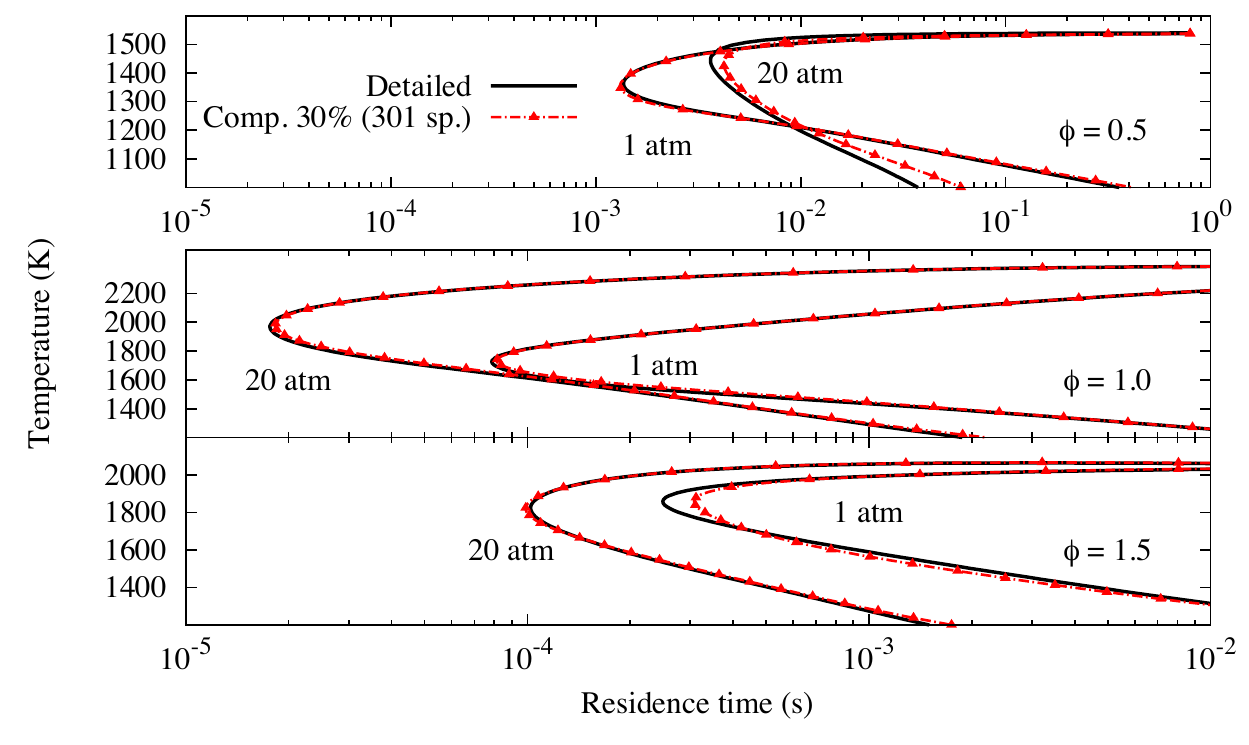}
\caption{Comparisons of PSR temperature response curves for the detailed TRF mechanism and the 30\%-error comprehensive skeletal mechanism generated using $\epsilon^* = 0.05$ for pressures of 1 and \SI{20}{\atm}, equivalence ratios of 0.5--1.5, and an inlet temperature of \SI{300}{\kelvin}.}
\label{F:comp-psr-05}
\end{figure}

\begin{figure}[htbp]
\centering
\includegraphics[width=\textwidth]{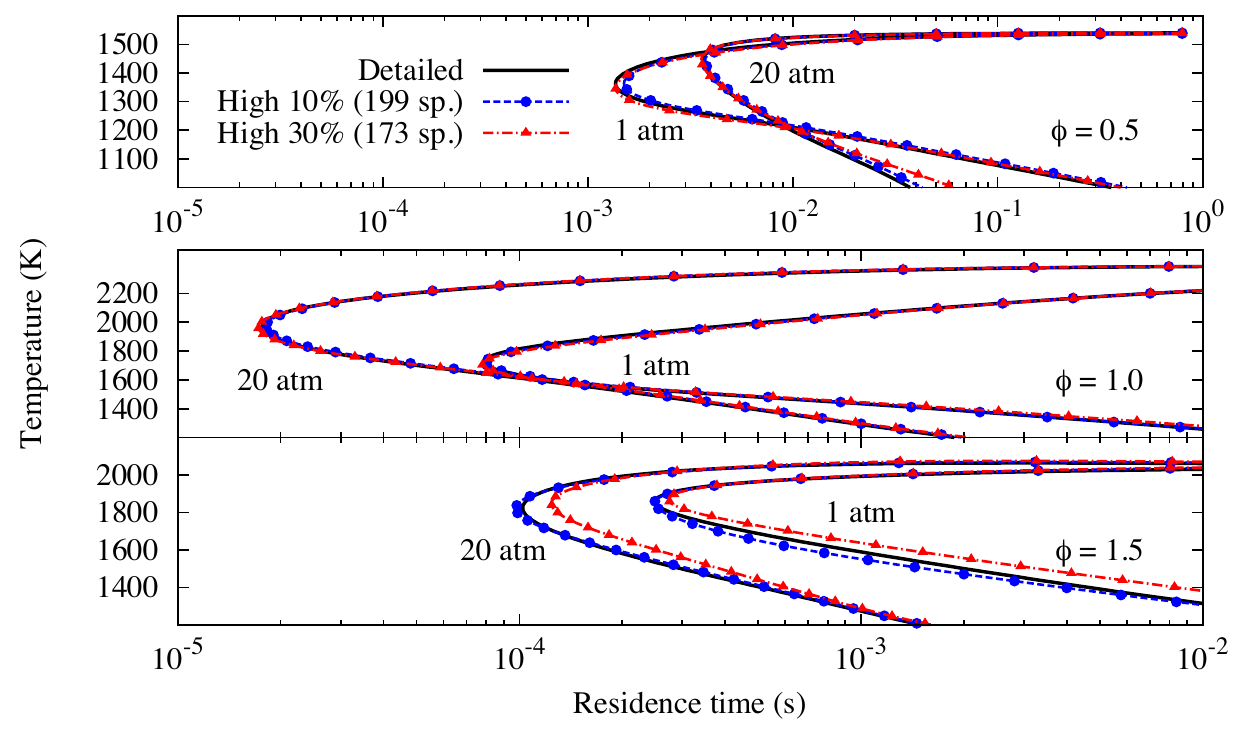}
\caption{Comparisons of PSR temperature response curves for the detailed TRF mechanism and two high-temperature skeletal mechanisms for pressures of 1 and \SI{20}{\atm}, equivalence ratios of 0.5--1.5, and an inlet temperature of \SI{300}{\kelvin}.}
\label{F:high-psr}
\end{figure}

\begin{figure}[htbp]
\centering
\includegraphics[width=\textwidth]{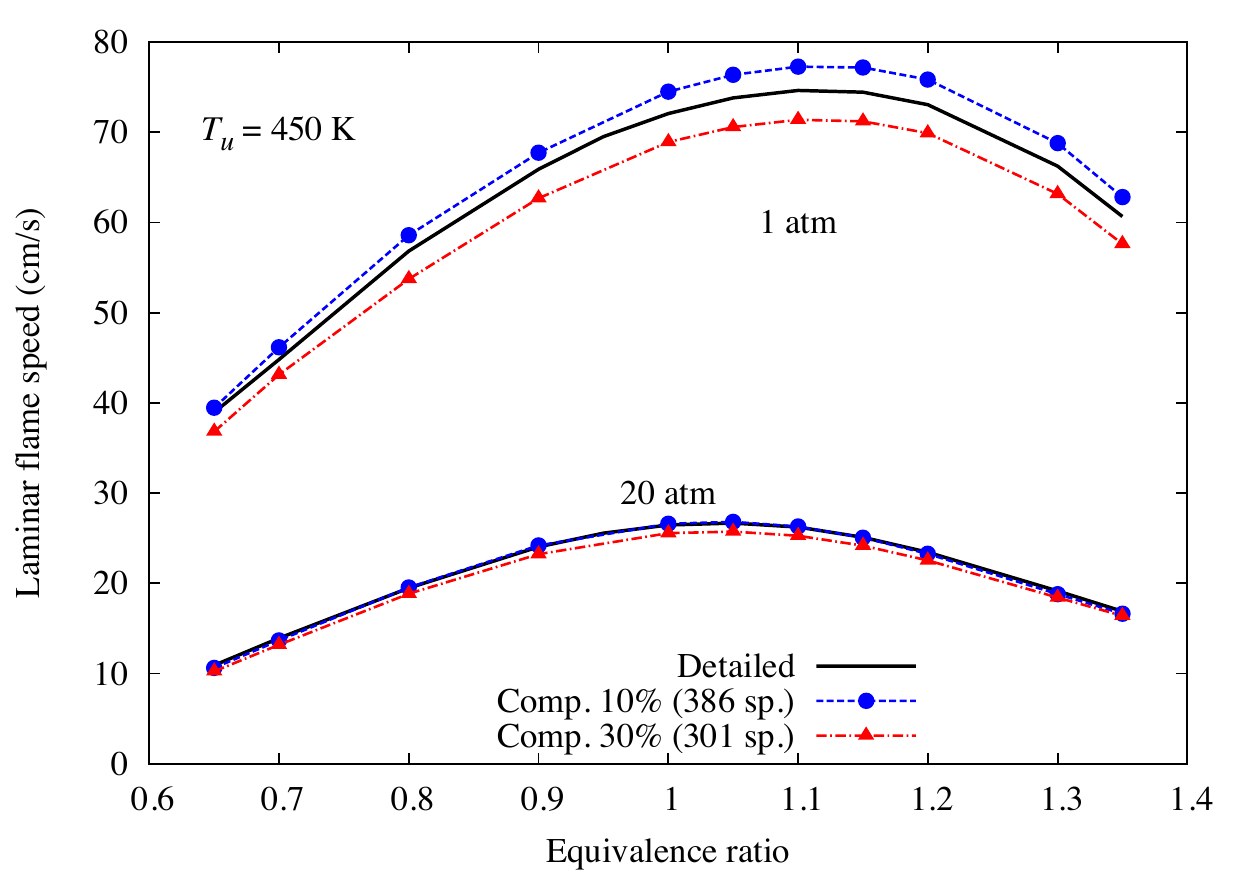}
\caption{Comparison of laminar flame speeds predicted using the detailed TRF mechanism and the two comprehensive skeletal mechanisms for 1 and \SI{20}{\atm} and an unburned gas temperature of $T_u$ = \SI{450}{\kelvin}, over a range of equivalence ratios. The 386 species 10\%-error skeletal mechanism was generated using $\epsilon^* = 0.1$ while the 30\%-error mechanism with 301 species used $\epsilon^* = 0.05$.}
\label{F:comp-flame-speed}
\end{figure}

\begin{figure}[htbp]
\centering
\includegraphics[width=\textwidth]{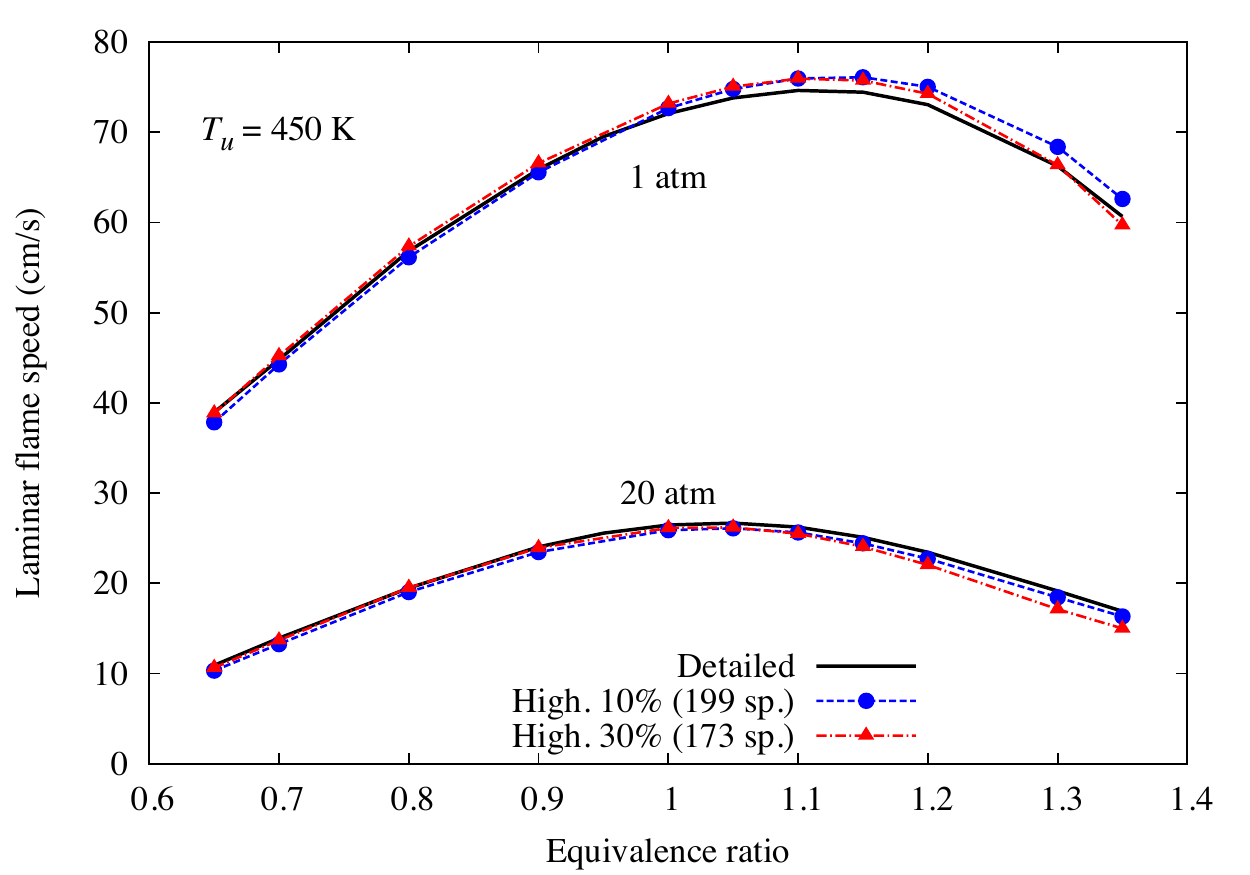}
\caption{Comparison of laminar flame speeds predicted using the detailed TRF mechanism and the 10\%- and 30\%-error high-temperature skeletal mechanisms for 1 and \SI{20}{\atm} and an unburned gas temperature of $T_u$ = \SI{450}{\kelvin}, over a range of equivalence ratios.}
\label{F:high-flame-speed}
\end{figure}

\begin{figure}[htbp]
\centering
\includegraphics[width=0.65\textwidth]{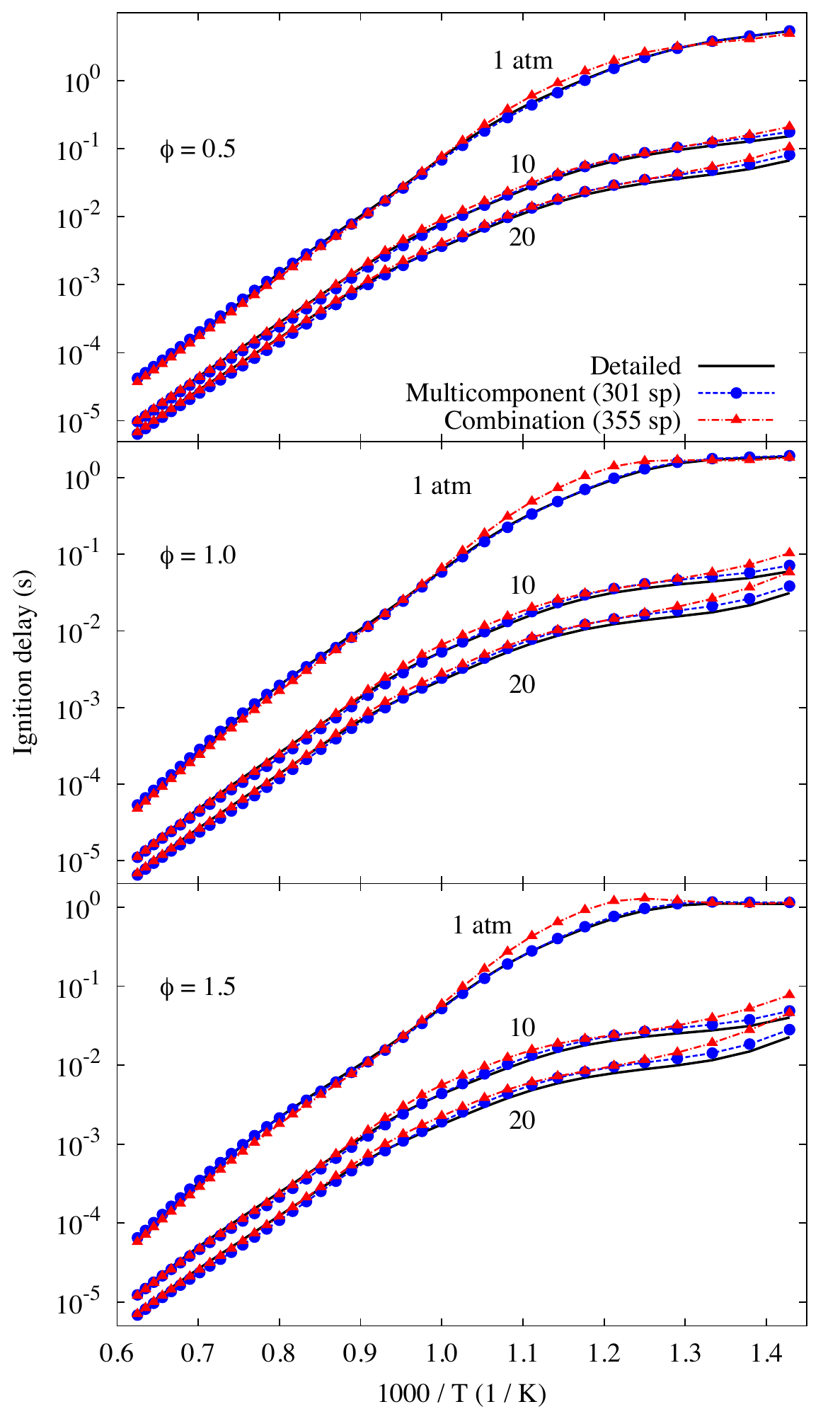}
\caption{Comparison of ignition delays calculated by multicomponent (301 species) and combination (355 species) skeletal mechanisms over a range of initial temperatures at various pressures and equivalence ratios.}
\label{F:combined-TRF-igndelay}
\end{figure}

\begin{figure}[htbp]
\centering
\includegraphics[width=.75\textwidth]{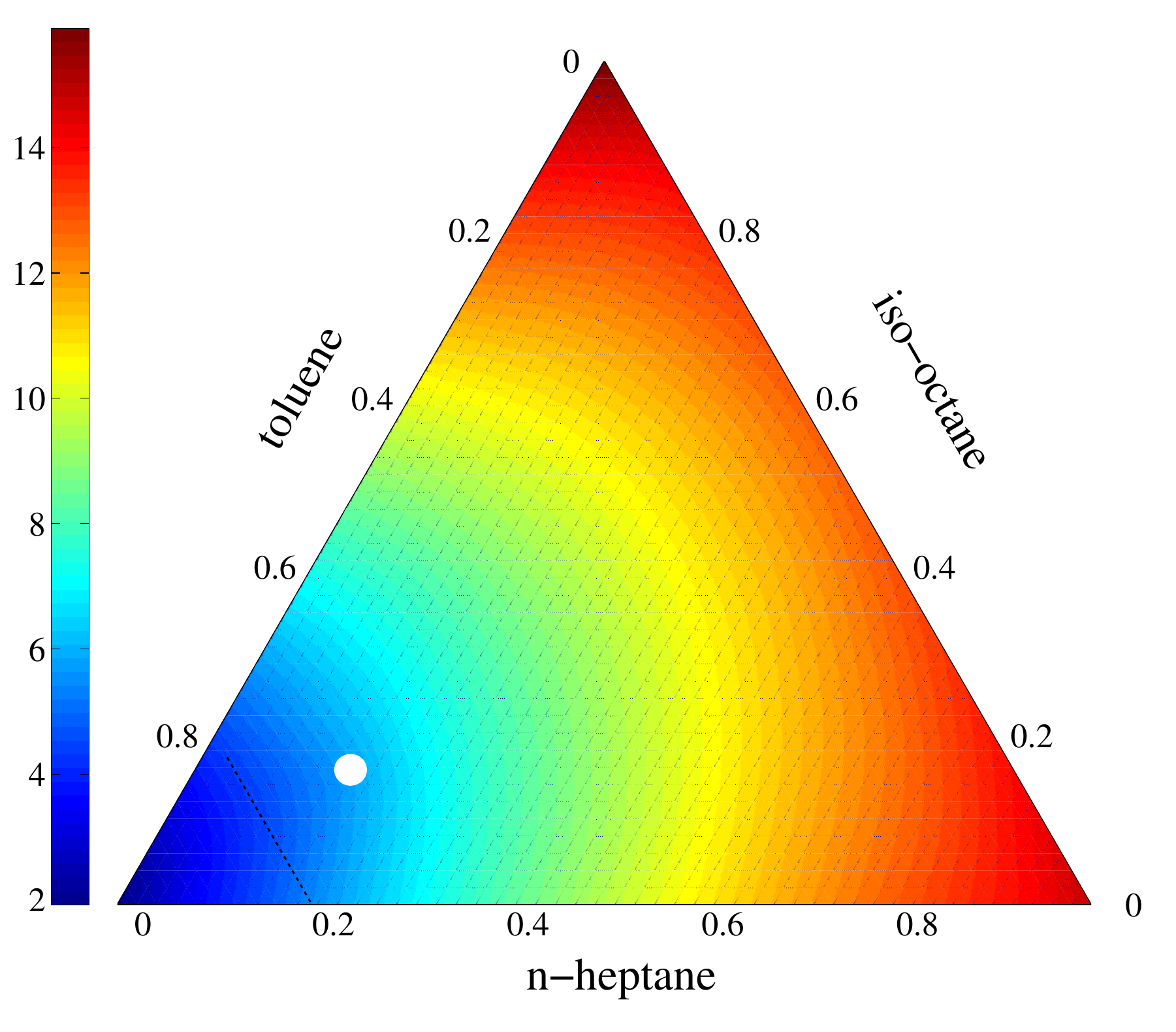}
\caption{Percent error in predicted autoignition delay for MON-like simulations using the 10\%-error comprehensive skeletal mechanism (386 species and \num{1591} reactions) with a varying fuel composition. The white circle indicates the TRF1 mixture composition used for the reduction procedure.}
\label{F:comp10-MON}
\end{figure}

\begin{figure}[htbp]
\centering
\includegraphics[width=.75\textwidth]{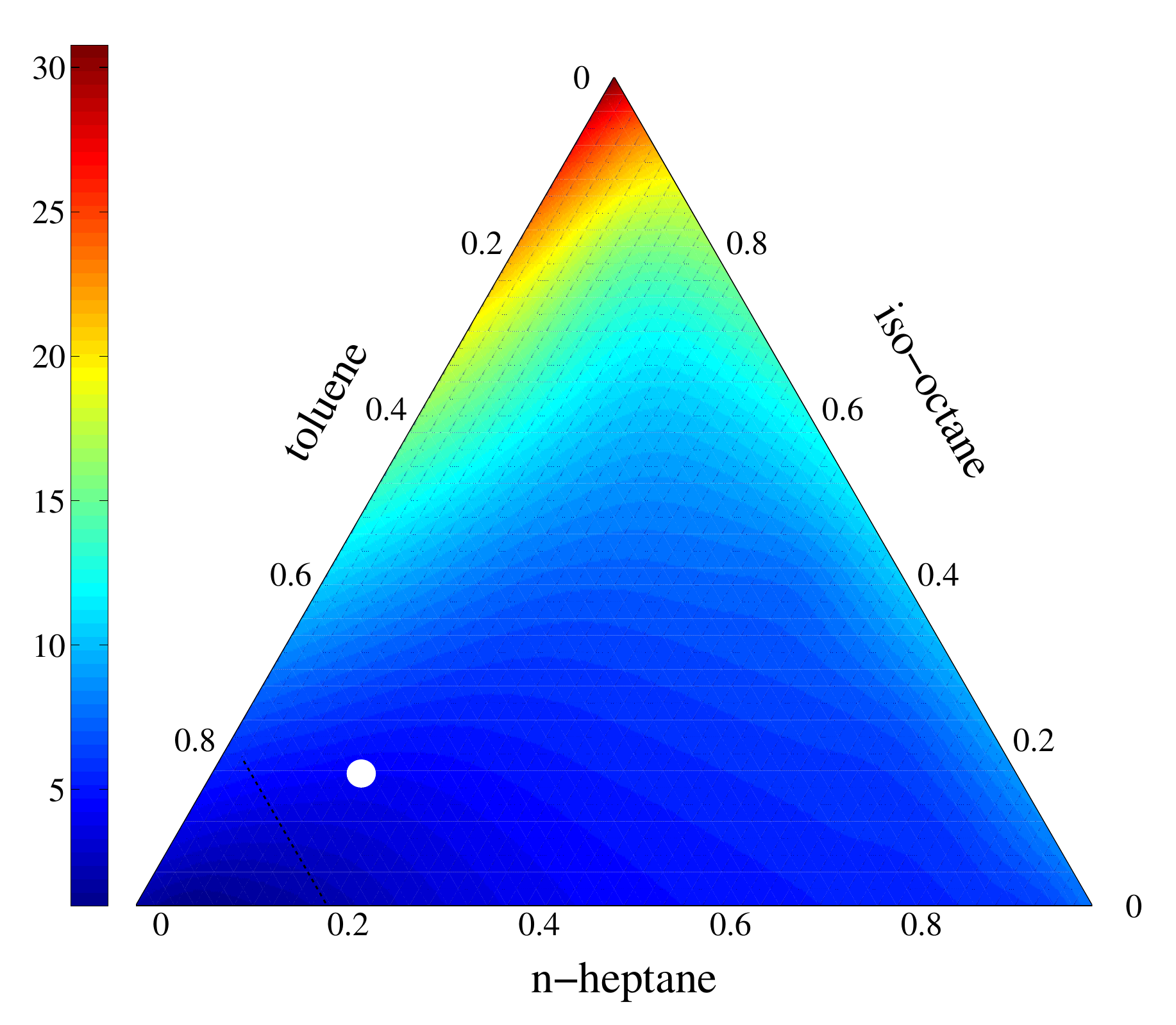}
\caption{Percent error in predicted autoignition delay for RON-like simulations using the 10\%-error comprehensive skeletal mechanism (386 species and \num{1591} reactions) with a varying fuel composition. The white circle indicates the TRF1 mixture composition used for the reduction procedure.}
\label{F:comp10-RON}
\end{figure}

\begin{figure}[htbp]
\centering
\includegraphics[width=.75\textwidth]{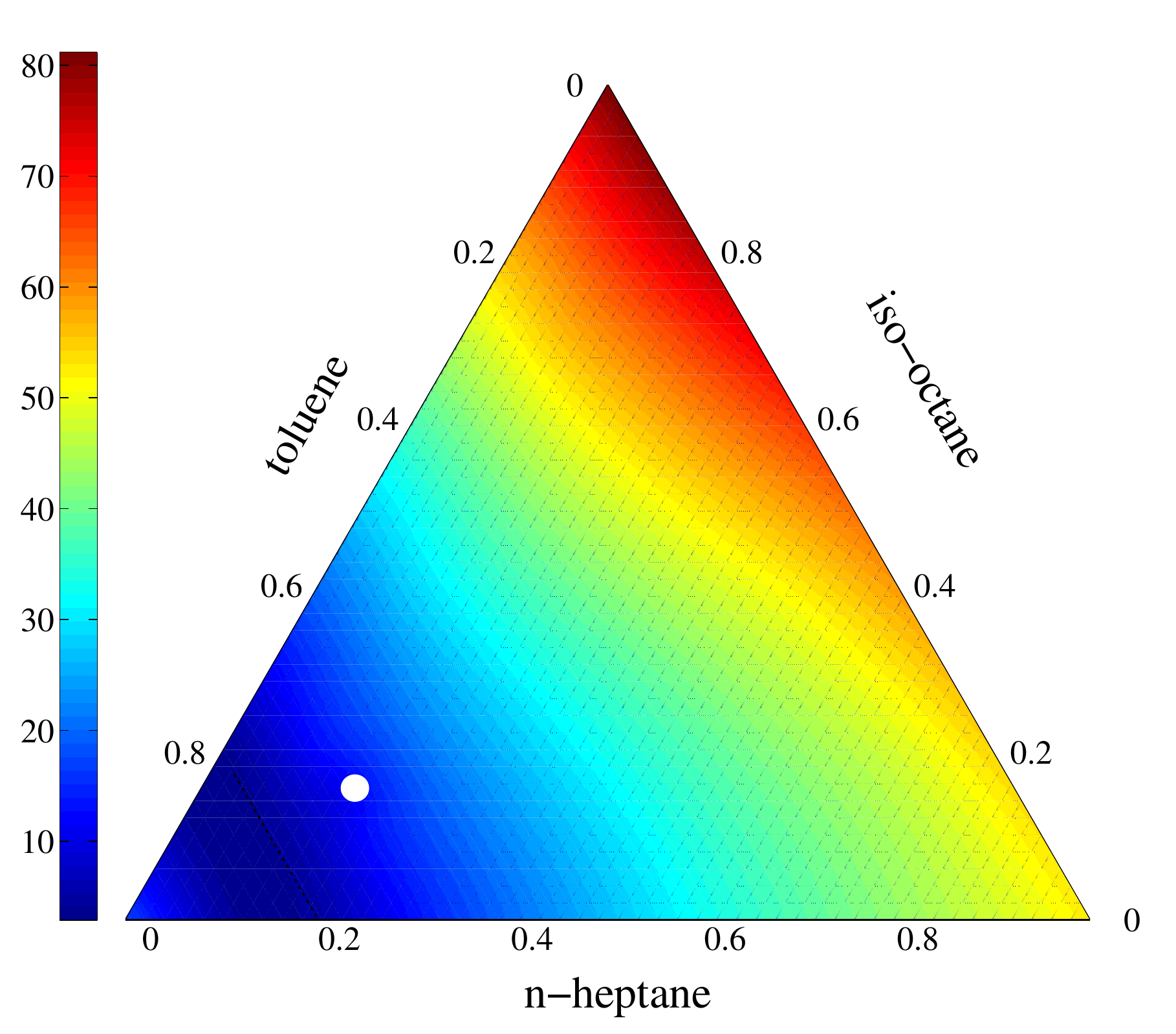}
\caption{Percent error in predicted autoignition delay for MON-like simulations using the 30\%-error comprehensive skeletal mechanism (276 species and 936 reactions) with a varying fuel composition. The white circle indicates the TRF1 mixture composition used for the reduction procedure.}
\label{F:comp30-MON}
\end{figure}

\begin{figure}[htbp]
\centering
\includegraphics[width=.75\textwidth]{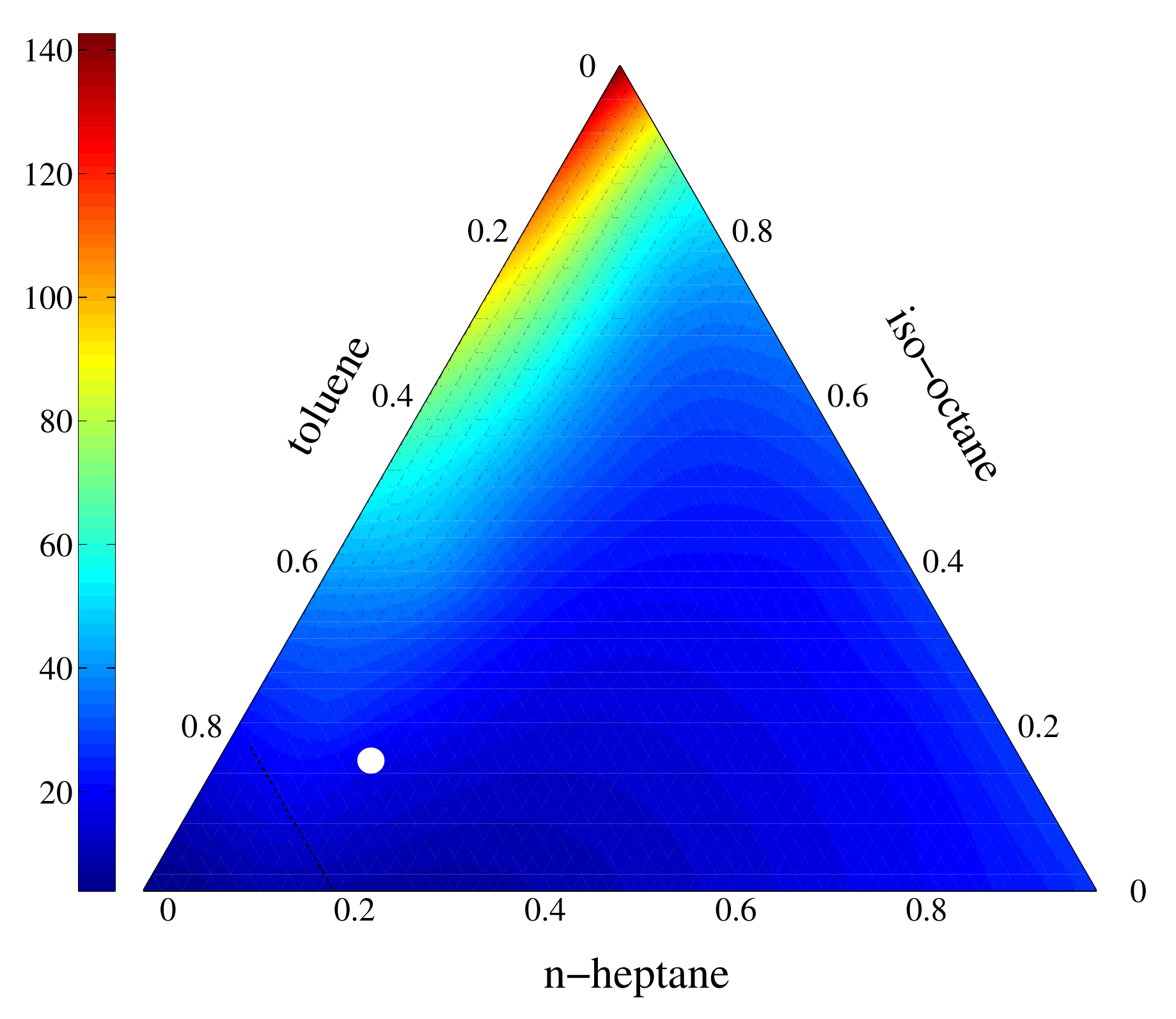}
\caption{Percent error in predicted autoignition delay for RON-like simulations using the 30\%-error comprehensive skeletal mechanism (276 species and 936 reactions) with a varying fuel composition. The white circle indicates the TRF1 mixture composition used for the reduction procedure.}
\label{F:comp30-RON}
\end{figure}

\begin{figure}[htbp]
\centering
\includegraphics[width=.75\textwidth]{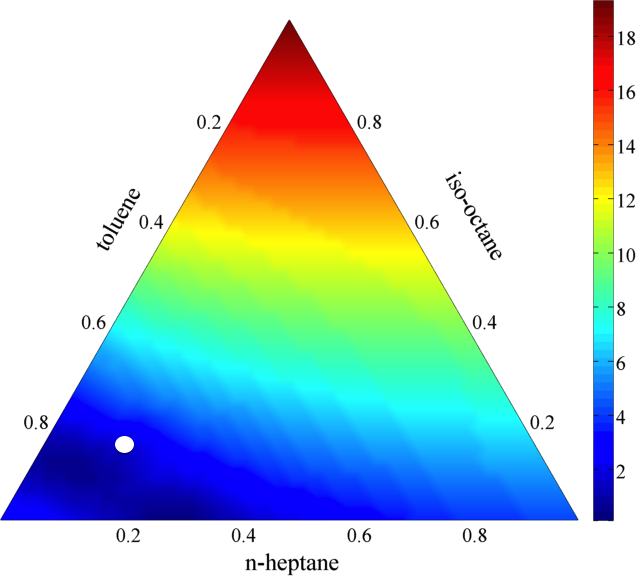}
\caption{Percent error in predicted ignition delay for constant volume autoignition using the 10\%-error high-temperature skeletal mechanism (199 species and \num{1011} reactions) with a varying fuel composition, using initial conditions of \SI{1200}{\kelvin}, \SI{10}{\atm}, and $\phi = 1.0$. The white circle indicates the TRF1 mixture composition used for the reduction procedure.}
\label{F:high10}
\end{figure}

\begin{figure}[htbp]
\centering
\begin{subfigure}[b]{0.8\textwidth}
	\includegraphics[width=\textwidth]{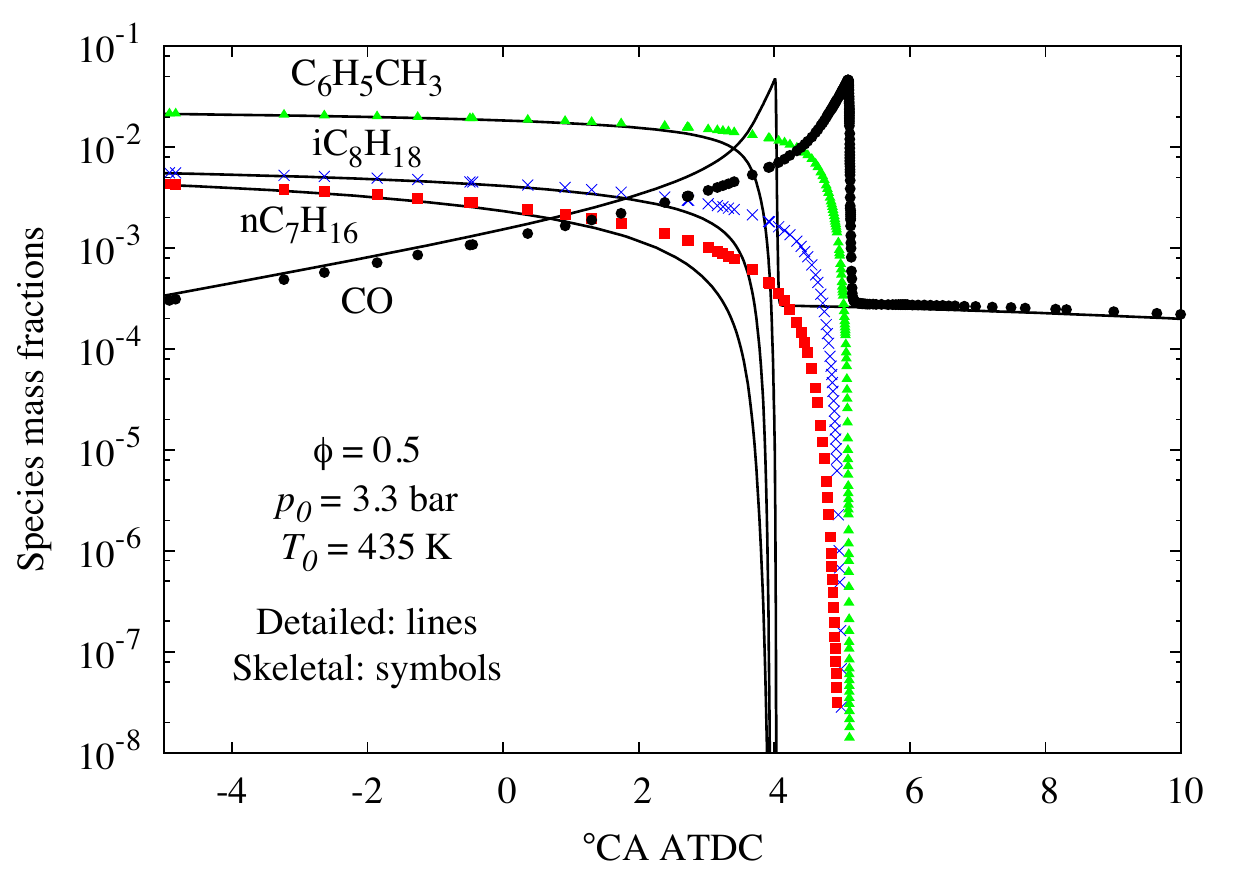}
	\caption{Original \textdegree CA}
\end{subfigure}

\begin{subfigure}[b]{0.8\textwidth}
	\includegraphics[width=\textwidth]{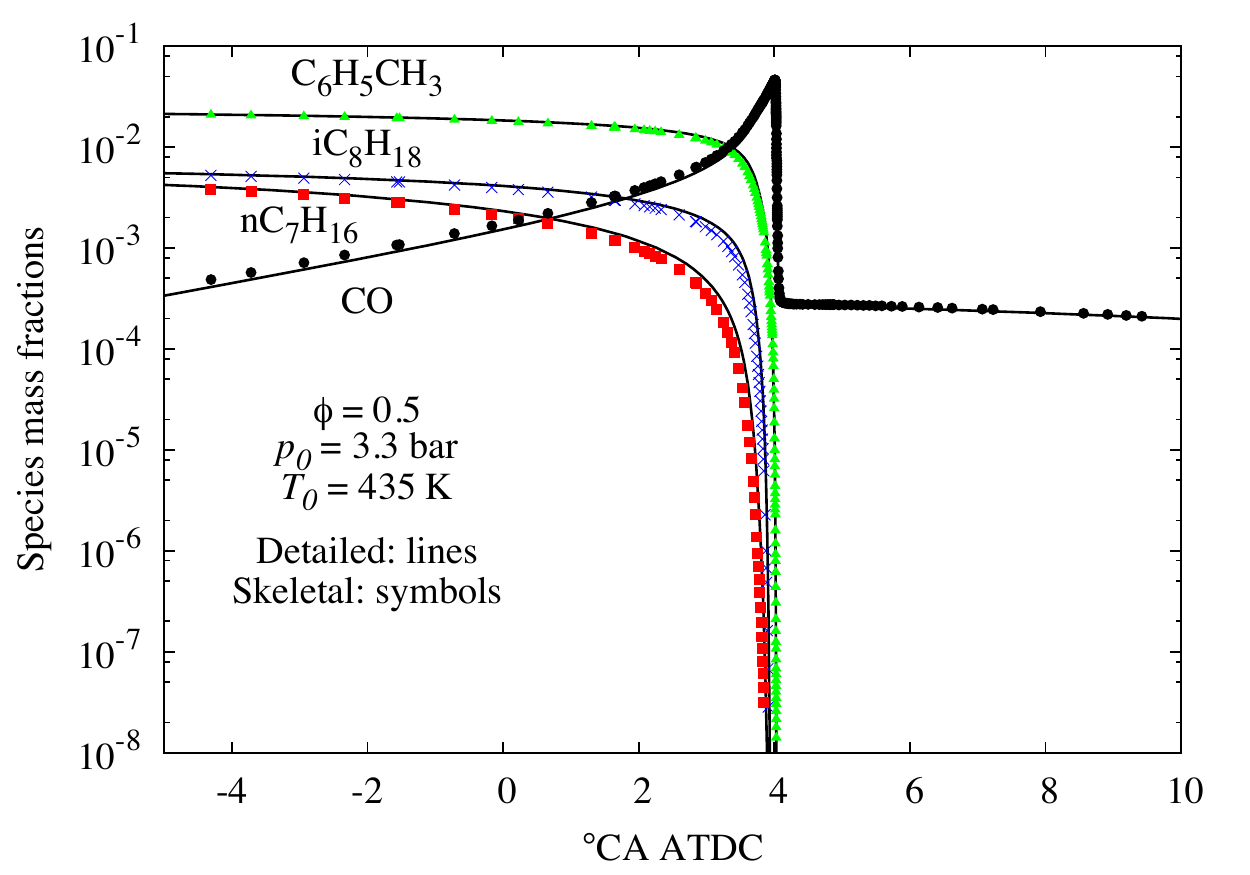}
	\caption{Shifted \textdegree CA}
\end{subfigure}
\caption{Comparison of major species mass fractions between the detailed TRF and 10\%-error comprehensive skeletal (386 species and \num{1591} reactions) mechanisms for a single-zone HCCI simulation with the IC5 initial conditions specified in Table~\ref{T:engine-IC}. The shifted \textdegree CA plot is based on ignition delay timing.}
\label{F:IC5-10}
\end{figure}

\begin{figure}[htbp]
\centering
\begin{subfigure}[b]{0.8\textwidth}
	\includegraphics[width=\textwidth]{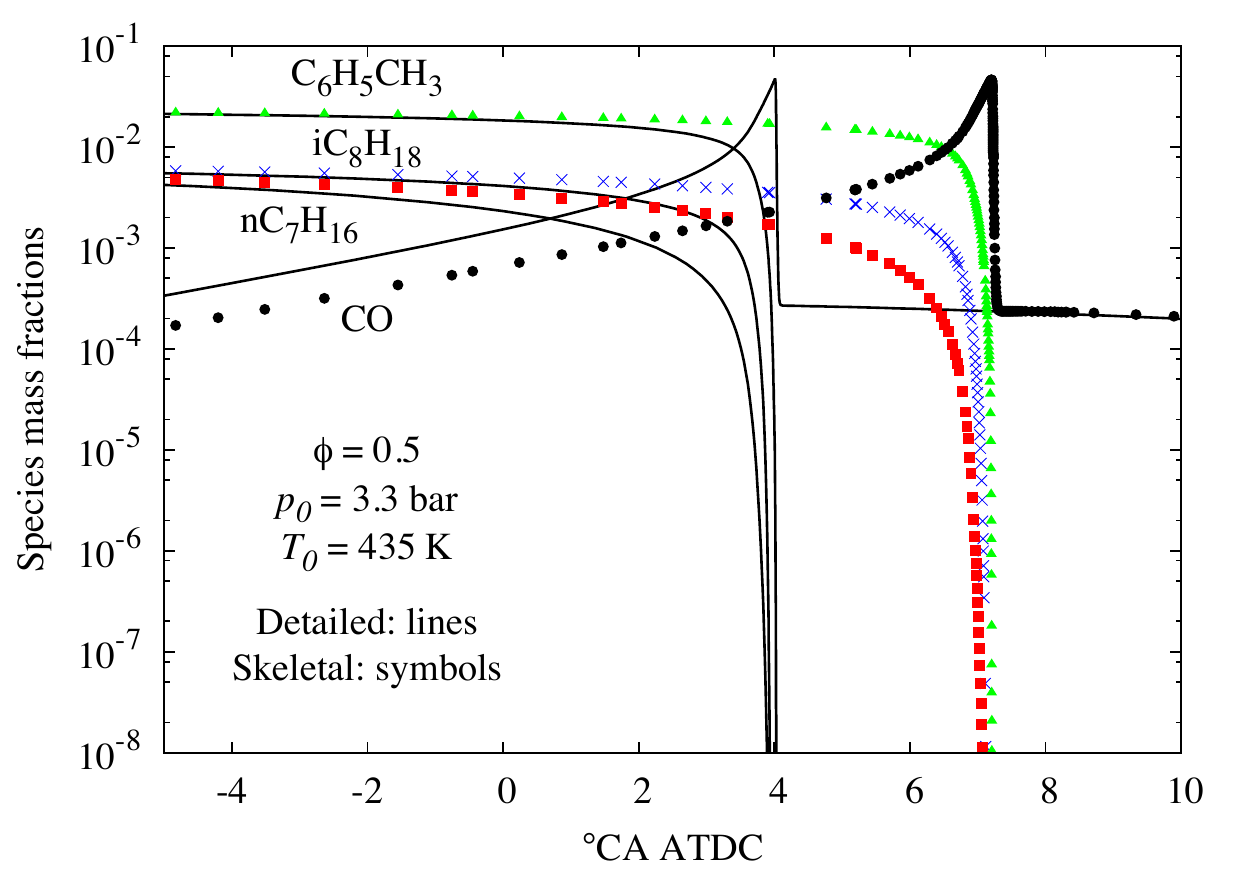}
	\caption{Original \textdegree CA}
\end{subfigure}

\begin{subfigure}[b]{0.8\textwidth}
	\includegraphics[width=\textwidth]{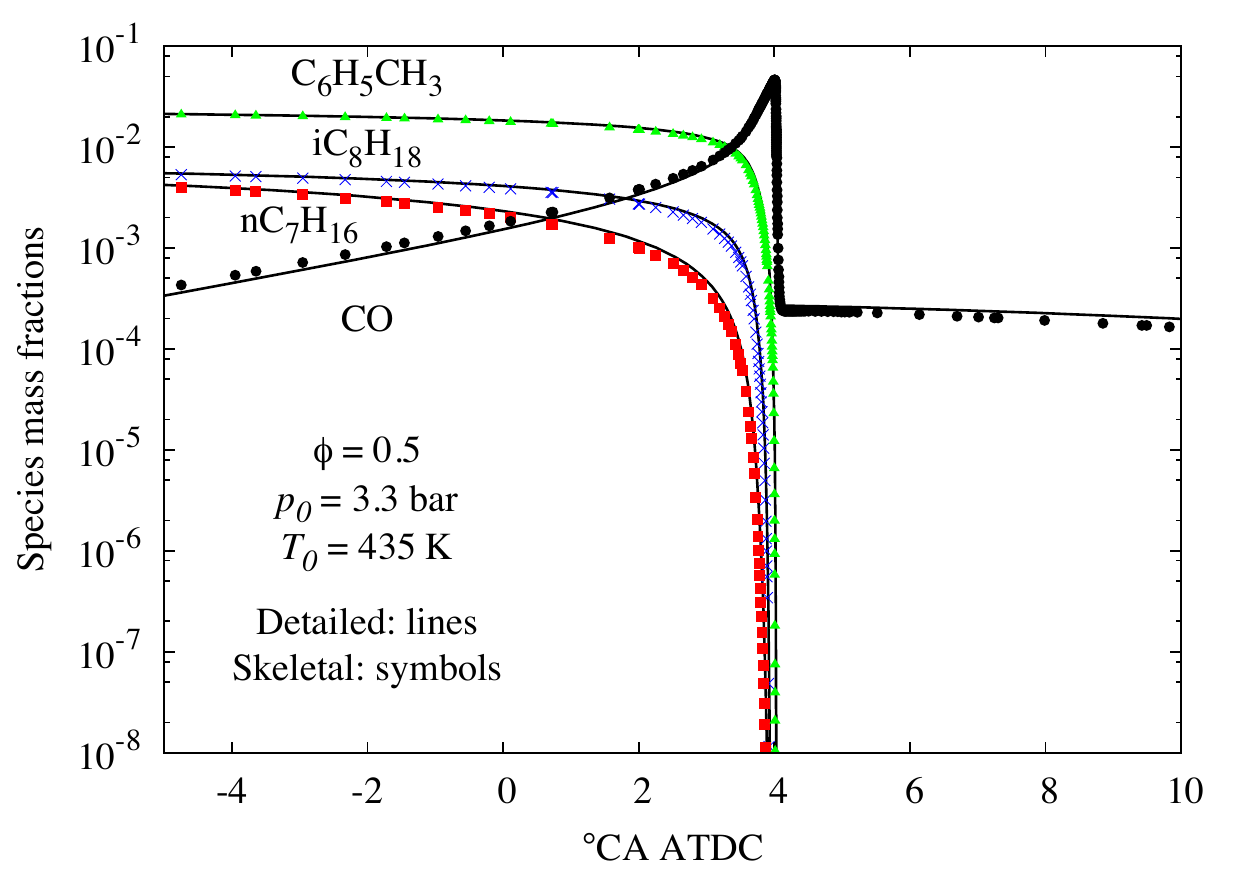}
	\caption{Shifted \textdegree CA}
\end{subfigure}
\caption{Comparison of major species mass fractions between the detailed TRF and 30\%-error comprehensive skeletal (276 species and 936 reactions) mechanisms for a single-zone HCCI simulation with the IC5 initial conditions specified in Table~\ref{T:engine-IC}. The shifted \textdegree CA plot is based on ignition delay timing.}
\label{F:IC5-30}
\end{figure}

\clearpage

%


\begin{thebibliography}{100}
\expandafter\ifx\csname natexlab\endcsname\relax\def\natexlab#1{#1}\fi
\providecommand{\bibinfo}[2]{#2}

\bibitem[{Edgar(1927)}]{Edgar:1927fq}
\bibinfo{author}{G.~Edgar}, \bibinfo{journal}{Ind. Eng. Chem.}
  \bibinfo{volume}{19} (\bibinfo{year}{1927}) \bibinfo{pages}{145--146}.

\bibitem[{Sturgis et~al.(1954)Sturgis, Cantwell, Morris, and
  Schultz}]{Sturgis:1954uq}
\bibinfo{author}{B.~M. Sturgis}, \bibinfo{author}{E.~N. Cantwell},
  \bibinfo{author}{W.~E. Morris}, \bibinfo{author}{D.~L. Schultz},
  \bibinfo{journal}{Proc. API} \bibinfo{volume}{34} (\bibinfo{year}{1954})
  \bibinfo{pages}{256--269}.

\bibitem[{Pahnke et~al.(1954)Pahnke, Cohen, and Sturgis}]{Pahnke:1954er}
\bibinfo{author}{A.~J. Pahnke}, \bibinfo{author}{P.~M. Cohen},
  \bibinfo{author}{B.~M. Sturgis}, \bibinfo{journal}{Ind. Eng. Chem.}
  \bibinfo{volume}{46} (\bibinfo{year}{1954}) \bibinfo{pages}{1024--1029}.

\bibitem[{{Universal Oil Products Co.}(1940)}]{Universal-Oil-Products-Co.:1940}
\bibinfo{author}{{Universal Oil Products Co.}}, \bibinfo{title}{Laboratory test
  methods for petroleum and its products}, \bibinfo{howpublished}{{Method No.
  H-173---40}}, \bibinfo{year}{1940}.

\bibitem[{Wei and Kuo(1969)}]{Wei:1969kt}
\bibinfo{author}{J.~Wei}, \bibinfo{author}{J.~C.~W. Kuo},
  \bibinfo{journal}{Ind. Eng. Chem. Fundam.} \bibinfo{volume}{8}
  (\bibinfo{year}{1969}) \bibinfo{pages}{114--123}.

\bibitem[{Coxson and Bischoff(1987)}]{Coxson:1987uj}
\bibinfo{author}{P.~G. Coxson}, \bibinfo{author}{K.~B. Bischoff},
  \bibinfo{journal}{Ind. Eng. Chem. Res.} \bibinfo{volume}{26}
  (\bibinfo{year}{1987}) \bibinfo{pages}{1239--1248}.

\bibitem[{Bailey(2001)}]{Bailey:2001gp}
\bibinfo{author}{J.~E. Bailey}, \bibinfo{journal}{Chem. Eng. J.}
  \bibinfo{volume}{3} (\bibinfo{year}{2001}) \bibinfo{pages}{52--61}.

\bibitem[{Lu and Law(2008)}]{Lu:2008bi}
\bibinfo{author}{T.~Lu}, \bibinfo{author}{C.~K. Law},
  \bibinfo{journal}{Combust. Flame} \bibinfo{volume}{154}
  (\bibinfo{year}{2008}) \bibinfo{pages}{153--163}.

\bibitem[{Pepiot and Pitsch(2008)}]{Pepiot:2008kq}
\bibinfo{author}{P.~Pepiot}, \bibinfo{author}{H.~Pitsch},
  \bibinfo{journal}{Combust. Theor. Model.} \bibinfo{volume}{12}
  (\bibinfo{year}{2008}) \bibinfo{pages}{1089--1108}.

\bibitem[{Curran et~al.(1998)Curran, Gaffuri, Pitz, and
  Westbrook}]{Curran:1998}
\bibinfo{author}{H.~Curran}, \bibinfo{author}{P.~Gaffuri},
  \bibinfo{author}{W.~Pitz}, \bibinfo{author}{C.~K. Westbrook},
  \bibinfo{journal}{Combust. Flame} \bibinfo{volume}{114}
  (\bibinfo{year}{1998}) \bibinfo{pages}{149--177}.

\bibitem[{Gauthier et~al.(2004)Gauthier, Davidson, and Hanson}]{Gauthier:2004}
\bibinfo{author}{B.~M. Gauthier}, \bibinfo{author}{D.~F. Davidson},
  \bibinfo{author}{R.~K. Hanson}, \bibinfo{journal}{Combust. Flame}
  \bibinfo{volume}{139} (\bibinfo{year}{2004}) \bibinfo{pages}{300--311}.

\bibitem[{Andrae et~al.(2007)Andrae, Bjornbom, Cracknell, and
  Kalghatgi}]{Andrae:2007}
\bibinfo{author}{J.~C.~G. Andrae}, \bibinfo{author}{P.~Bjornbom},
  \bibinfo{author}{R.~Cracknell}, \bibinfo{author}{G.~Kalghatgi},
  \bibinfo{journal}{Combust. Flame} \bibinfo{volume}{149}
  (\bibinfo{year}{2007}) \bibinfo{pages}{2--24}.

\bibitem[{Chaos et~al.(2007)Chaos, Zhao, Kazakov, Gokulakrishnan, Angioletti,
  and Dryer}]{Chaos:2007}
\bibinfo{author}{M.~Chaos}, \bibinfo{author}{Z.~Zhao},
  \bibinfo{author}{A.~Kazakov}, \bibinfo{author}{P.~Gokulakrishnan},
  \bibinfo{author}{M.~Angioletti}, \bibinfo{author}{F.~Dryer}, in:
  \bibinfo{booktitle}{5th US Combustion Meeting}, \bibinfo{number}{E-62}.

\bibitem[{Andrae et~al.(2008)Andrae, Brinck, and Kalghatgi}]{Andrae:2008}
\bibinfo{author}{J.~C.~G. Andrae}, \bibinfo{author}{T.~Brinck},
  \bibinfo{author}{G.~Kalghatgi}, \bibinfo{journal}{Combust. Flame}
  \bibinfo{volume}{155} (\bibinfo{year}{2008}) \bibinfo{pages}{696--712}.

\bibitem[{Liang et~al.(2009)Liang, Stevens, Raman, and Farrell}]{Liang:2009a}
\bibinfo{author}{L.~Liang}, \bibinfo{author}{J.~G. Stevens},
  \bibinfo{author}{S.~Raman}, \bibinfo{author}{J.~T. Farrell},
  \bibinfo{journal}{Combust. Flame} \bibinfo{volume}{156}
  (\bibinfo{year}{2009}) \bibinfo{pages}{1493--1502}.

\bibitem[{Morgan et~al.(2010)Morgan, Smallbone, Bhave, Kraft, Cracknell, and
  Kalghatgi}]{Morgan:2010}
\bibinfo{author}{N.~Morgan}, \bibinfo{author}{A.~Smallbone},
  \bibinfo{author}{A.~Bhave}, \bibinfo{author}{M.~Kraft},
  \bibinfo{author}{R.~Cracknell}, \bibinfo{author}{G.~Kalghatgi},
  \bibinfo{journal}{Combust. Flame} \bibinfo{volume}{157}
  (\bibinfo{year}{2010}) \bibinfo{pages}{1122--1131}.

\bibitem[{Kumar and Sung(2010)}]{Kumar:2010js}
\bibinfo{author}{K.~Kumar}, \bibinfo{author}{C.~J. Sung},
  \bibinfo{journal}{Energy Fuels} \bibinfo{volume}{24} (\bibinfo{year}{2010})
  \bibinfo{pages}{3840--3849}.

\bibitem[{Hartmann et~al.(2011)Hartmann, Gushterova, Fikri, Schulz,
  Schie{\ss}l, and Maas}]{Hartmann:2011ku}
\bibinfo{author}{M.~Hartmann}, \bibinfo{author}{I.~Gushterova},
  \bibinfo{author}{M.~Fikri}, \bibinfo{author}{C.~Schulz},
  \bibinfo{author}{R.~Schie{\ss}l}, \bibinfo{author}{U.~Maas},
  \bibinfo{journal}{Combust. Flame} \bibinfo{volume}{158}
  (\bibinfo{year}{2011}) \bibinfo{pages}{172--178}.

\bibitem[{Knop et~al.(2013)Knop, Pera, and Duffour}]{Knop:2013gt}
\bibinfo{author}{V.~Knop}, \bibinfo{author}{C.~Pera},
  \bibinfo{author}{F.~Duffour}, \bibinfo{journal}{Combust. Flame}
  \bibinfo{volume}{160} (\bibinfo{year}{2013}) \bibinfo{pages}{2067--2082}.

\bibitem[{Kukkadapu et~al.(2013)Kukkadapu, Kumar, Sung, Mehl, and
  Pitz}]{Kukkadapu:2013ko}
\bibinfo{author}{G.~Kukkadapu}, \bibinfo{author}{K.~Kumar},
  \bibinfo{author}{C.~J. Sung}, \bibinfo{author}{M.~Mehl},
  \bibinfo{author}{W.~J. Pitz}, \bibinfo{journal}{Proc. Combust. Inst.}
  \bibinfo{volume}{34} (\bibinfo{year}{2013}) \bibinfo{pages}{345--352}.

\bibitem[{Rapp et~al.(2013)Rapp, Cannella, Chen, and Dibble}]{Rapp:2013ev}
\bibinfo{author}{V.~H. Rapp}, \bibinfo{author}{W.~J. Cannella},
  \bibinfo{author}{J.-Y. Chen}, \bibinfo{author}{R.~W. Dibble},
  \bibinfo{journal}{Combust. Sci. Technol.} \bibinfo{volume}{185}
  (\bibinfo{year}{2013}) \bibinfo{pages}{735--748}.

\bibitem[{Pitz et~al.(2007)Pitz, Cernansky, Dryer, Egolfopoulos, Farrell,
  Friend, and Pitsch}]{Pitz:2007}
\bibinfo{author}{W.~J. Pitz}, \bibinfo{author}{N.~P. Cernansky},
  \bibinfo{author}{F.~Dryer}, \bibinfo{author}{F.~Egolfopoulos},
  \bibinfo{author}{J.~T. Farrell}, \bibinfo{author}{D.~Friend},
  \bibinfo{author}{H.~Pitsch}, \bibinfo{title}{Development of an experimental
  database and chemical kinetic models for surrogate gasoline fuels},
  \bibinfo{howpublished}{SAE 2007-01-0175}, \bibinfo{year}{2007}.

\bibitem[{Farrell et~al.(2007)Farrell, Cernansky, Dryer, Friend, Hergart, Law,
  McDavid, Mueller, Patel, and Pitsch}]{Farrell:2007}
\bibinfo{author}{J.~T. Farrell}, \bibinfo{author}{N.~P. Cernansky},
  \bibinfo{author}{F.~Dryer}, \bibinfo{author}{D.~Friend},
  \bibinfo{author}{C.~A. Hergart}, \bibinfo{author}{C.~K. Law},
  \bibinfo{author}{R.~M. McDavid}, \bibinfo{author}{C.~J. Mueller},
  \bibinfo{author}{A.~K. Patel}, \bibinfo{author}{H.~Pitsch},
  \bibinfo{title}{Development of an experimental database and kinetic models
  for surrogate diesel fuels}, \bibinfo{howpublished}{SAE 2007-01-0201},
  \bibinfo{year}{2007}.

\bibitem[{Pitz and Mueller(2011)}]{Pitz:2011iv}
\bibinfo{author}{W.~J. Pitz}, \bibinfo{author}{C.~J. Mueller},
  \bibinfo{journal}{Prog. Energy Comb. Sci.} \bibinfo{volume}{37}
  (\bibinfo{year}{2011}) \bibinfo{pages}{330--350}.

\bibitem[{Colket et~al.(2007)Colket, Edwards, Williams, Cernansky, Miller,
  Egolfopoulos, Lindstedt, Seshadri, Dryer, Law, Friend, Lenhert, Pitsch,
  Sarofim, Smooke, and Tsang}]{Colket:2007}
\bibinfo{author}{M.~Colket}, \bibinfo{author}{T.~Edwards},
  \bibinfo{author}{S.~Williams}, \bibinfo{author}{N.~P. Cernansky},
  \bibinfo{author}{D.~L. Miller}, \bibinfo{author}{F.~Egolfopoulos},
  \bibinfo{author}{P.~Lindstedt}, \bibinfo{author}{K.~Seshadri},
  \bibinfo{author}{F.~Dryer}, \bibinfo{author}{C.~K. Law},
  \bibinfo{author}{D.~Friend}, \bibinfo{author}{D.~B. Lenhert},
  \bibinfo{author}{H.~Pitsch}, \bibinfo{author}{A.~F. Sarofim},
  \bibinfo{author}{M.~Smooke}, \bibinfo{author}{W.~Tsang}, in:
  \bibinfo{booktitle}{45th AIAA Aerospace Sciences Meeting},
  \bibinfo{number}{AIAA 2007-770, 2007}.

\bibitem[{Colket et~al.(2008)Colket, Edwards, Williams, Cernansky, Miller,
  Egolfopoulos, Dryer, Bellan, Lindstedt, Seshadri, Pitsch, Sarofim, Smooke,
  and Tsang}]{Colket:2008}
\bibinfo{author}{M.~Colket}, \bibinfo{author}{T.~Edwards},
  \bibinfo{author}{S.~Williams}, \bibinfo{author}{N.~P. Cernansky},
  \bibinfo{author}{D.~L. Miller}, \bibinfo{author}{F.~Egolfopoulos},
  \bibinfo{author}{F.~Dryer}, \bibinfo{author}{J.~Bellan},
  \bibinfo{author}{P.~Lindstedt}, \bibinfo{author}{K.~Seshadri},
  \bibinfo{author}{H.~Pitsch}, \bibinfo{author}{A.~Sarofim},
  \bibinfo{author}{M.~Smooke}, \bibinfo{author}{W.~Tsang}, in:
  \bibinfo{booktitle}{46th AIAA Aerospace Sciences Meeting},
  \bibinfo{number}{AIAA 2008-972, 2008}.

\bibitem[{Mehl et~al.(2011)Mehl, Pitz, Westbrook, and Curran}]{Mehl:2011cn}
\bibinfo{author}{M.~Mehl}, \bibinfo{author}{W.~J. Pitz}, \bibinfo{author}{C.~K.
  Westbrook}, \bibinfo{author}{H.~J. Curran}, \bibinfo{journal}{Proc. Combust.
  Inst.} \bibinfo{volume}{33} (\bibinfo{year}{2011}) \bibinfo{pages}{193--200}.

\bibitem[{Lenhert et~al.(2009)Lenhert, Miller, Cernansky, and
  Owens}]{Lenhert:2009}
\bibinfo{author}{D.~B. Lenhert}, \bibinfo{author}{D.~L. Miller},
  \bibinfo{author}{N.~P. Cernansky}, \bibinfo{author}{K.~G. Owens},
  \bibinfo{journal}{Combust. Flame} \bibinfo{volume}{156}
  (\bibinfo{year}{2009}) \bibinfo{pages}{549--564}.

\bibitem[{Mehl et~al.(2011)Mehl, Chen, Pitz, Sarathy, and
  Westbrook}]{Mehl:2011jn}
\bibinfo{author}{M.~Mehl}, \bibinfo{author}{J.-Y. Chen}, \bibinfo{author}{W.~J.
  Pitz}, \bibinfo{author}{S.~M. Sarathy}, \bibinfo{author}{C.~K. Westbrook},
  \bibinfo{journal}{Energy Fuels} \bibinfo{volume}{25} (\bibinfo{year}{2011})
  \bibinfo{pages}{5215--5223}.

\bibitem[{Naik et~al.(2005)Naik, Pitz, Westbrook, Sj{\"o}berg, Dec, Orme,
  Curran, and Simmie}]{Naik:2005}
\bibinfo{author}{C.~V. Naik}, \bibinfo{author}{W.~J. Pitz},
  \bibinfo{author}{C.~K. Westbrook}, \bibinfo{author}{M.~Sj{\"o}berg},
  \bibinfo{author}{J.~E. Dec}, \bibinfo{author}{J.~Orme},
  \bibinfo{author}{H.~J. Curran}, \bibinfo{author}{J.~M. Simmie},
  \bibinfo{title}{Detailed chemical kinetic modeling of surrogate fuels for
  gasoline and application to an {HCCI} engine}, \bibinfo{howpublished}{SAE
  Technical Paper 2005-01-3741}, \bibinfo{year}{2005}.

\bibitem[{Puduppakkam et~al.(2011)Puduppakkam, Liang, Naik, Meeks, Kokjohn, and
  Reitz}]{Puduppakkam:2011}
\bibinfo{author}{K.~V. Puduppakkam}, \bibinfo{author}{L.~Liang},
  \bibinfo{author}{C.~V. Naik}, \bibinfo{author}{E.~Meeks},
  \bibinfo{author}{S.~L. Kokjohn}, \bibinfo{author}{R.~D. Reitz},
  \bibinfo{journal}{SAE Int. J. Engines} \bibinfo{volume}{4}
  (\bibinfo{year}{2011}) \bibinfo{pages}{1127--1149}.

\bibitem[{Hentschel et~al.(1994)Hentschel, Schindler, and
  Haahtela}]{Hentschel:1994}
\bibinfo{author}{W.~Hentschel}, \bibinfo{author}{K.-P. Schindler},
  \bibinfo{author}{O.~Haahtela}, \bibinfo{title}{European diesel research
  {IDEA} --- experimental results from {DI} diesel engine investigations},
  \bibinfo{type}{Technical Report} \bibinfo{number}{SAE Technical Paper
  941954}, SAE International, \bibinfo{year}{1994}.

\bibitem[{Hernandez et~al.(2008)Hernandez, Sanz-Argent, Benajes, and
  Molina}]{Hernandez:2008bv}
\bibinfo{author}{J.~J. Hernandez}, \bibinfo{author}{J.~Sanz-Argent},
  \bibinfo{author}{J.~Benajes}, \bibinfo{author}{S.~Molina},
  \bibinfo{journal}{Fuel} \bibinfo{volume}{87} (\bibinfo{year}{2008})
  \bibinfo{pages}{655--665}.

\bibitem[{Mathieu et~al.(2009)Mathieu, Djeba{\"\i}li-Chaumeix, Paillard, and
  Douce}]{Mathieu:2009he}
\bibinfo{author}{O.~Mathieu}, \bibinfo{author}{N.~Djeba{\"\i}li-Chaumeix},
  \bibinfo{author}{C.-E. Paillard}, \bibinfo{author}{F.~Douce},
  \bibinfo{journal}{Combust. Flame} \bibinfo{volume}{156}
  (\bibinfo{year}{2009}) \bibinfo{pages}{1576--1586}.

\bibitem[{Mati et~al.(2007)Mati, Ristori, Ga{\"\i}l, Pengloan, and
  Dagaut}]{Mati:2007hh}
\bibinfo{author}{K.~Mati}, \bibinfo{author}{A.~Ristori},
  \bibinfo{author}{S.~Ga{\"\i}l}, \bibinfo{author}{G.~Pengloan},
  \bibinfo{author}{P.~Dagaut}, \bibinfo{journal}{Proc. Combust. Inst.}
  \bibinfo{volume}{31} (\bibinfo{year}{2007}) \bibinfo{pages}{2939--2946}.

\bibitem[{Mueller et~al.(2012)Mueller, Cannella, Bruno, Bunting, Dettman,
  Franz, Huber, Natarajan, Pitz, Ratcliff, and Wright}]{Mueller:2012gx}
\bibinfo{author}{C.~J. Mueller}, \bibinfo{author}{W.~J. Cannella},
  \bibinfo{author}{T.~J. Bruno}, \bibinfo{author}{B.~Bunting},
  \bibinfo{author}{H.~D. Dettman}, \bibinfo{author}{J.~A. Franz},
  \bibinfo{author}{M.~L. Huber}, \bibinfo{author}{M.~Natarajan},
  \bibinfo{author}{W.~J. Pitz}, \bibinfo{author}{M.~A. Ratcliff},
  \bibinfo{author}{K.~Wright}, \bibinfo{journal}{Energy Fuels}
  \bibinfo{volume}{26} (\bibinfo{year}{2012}) \bibinfo{pages}{3284--3303}.

\bibitem[{Krishnasamy et~al.(2013)Krishnasamy, Reitz, Willems, and
  Kurtz}]{Krishnasamy:2013bx}
\bibinfo{author}{A.~Krishnasamy}, \bibinfo{author}{R.~D. Reitz},
  \bibinfo{author}{W.~Willems}, \bibinfo{author}{E.~Kurtz},
  \bibinfo{title}{Surrogate Diesel Fuel Models for Low Temperature Combustion},
  \bibinfo{type}{Technical Report} \bibinfo{number}{SAE Technical Paper
  2013-01-1092}, SAE International, \bibinfo{year}{2013}.

\bibitem[{Herbinet et~al.(2008)Herbinet, Pitz, and Westbrook}]{Herbinet:2008iw}
\bibinfo{author}{O.~Herbinet}, \bibinfo{author}{W.~J. Pitz},
  \bibinfo{author}{C.~K. Westbrook}, \bibinfo{journal}{Combust. Flame}
  \bibinfo{volume}{154} (\bibinfo{year}{2008}) \bibinfo{pages}{507--528}.

\bibitem[{Liu et~al.(2013)Liu, Sivaramakrishnan, Davis, Som, Longman, and
  Lu}]{Liu:2013em}
\bibinfo{author}{W.~Liu}, \bibinfo{author}{R.~Sivaramakrishnan},
  \bibinfo{author}{M.~J. Davis}, \bibinfo{author}{S.~K. Som},
  \bibinfo{author}{D.~E. Longman}, \bibinfo{author}{T.~F. Lu},
  \bibinfo{journal}{Proc. Combust. Inst.} \bibinfo{volume}{34}
  (\bibinfo{year}{2013}) \bibinfo{pages}{401--409}.

\bibitem[{Herbinet et~al.(2010)Herbinet, Pitz, and Westbrook}]{Herbinet:2010}
\bibinfo{author}{O.~Herbinet}, \bibinfo{author}{W.~J. Pitz},
  \bibinfo{author}{C.~K. Westbrook}, \bibinfo{journal}{Combust. Flame}
  \bibinfo{volume}{157} (\bibinfo{year}{2010}) \bibinfo{pages}{893--908}.

\bibitem[{Honnet et~al.(2009)Honnet, Seshadri, Niemann, and
  Peters}]{Honnet:2009kl}
\bibinfo{author}{S.~Honnet}, \bibinfo{author}{K.~Seshadri},
  \bibinfo{author}{U.~Niemann}, \bibinfo{author}{N.~Peters},
  \bibinfo{journal}{Proc. Combust. Inst.} \bibinfo{volume}{32}
  (\bibinfo{year}{2009}) \bibinfo{pages}{485--492}.

\bibitem[{Dooley et~al.(2010)Dooley, Won, Chaos, Heyne, Ju, Dryer, Kumar, Sung,
  Wang, Oehlschlaeger, Santoro, and Litzinger}]{Dooley:2010js}
\bibinfo{author}{S.~Dooley}, \bibinfo{author}{S.~H. Won},
  \bibinfo{author}{M.~Chaos}, \bibinfo{author}{J.~Heyne},
  \bibinfo{author}{Y.~Ju}, \bibinfo{author}{F.~L. Dryer},
  \bibinfo{author}{K.~Kumar}, \bibinfo{author}{C.~J. Sung},
  \bibinfo{author}{H.~Wang}, \bibinfo{author}{M.~A. Oehlschlaeger},
  \bibinfo{author}{R.~J. Santoro}, \bibinfo{author}{T.~A. Litzinger},
  \bibinfo{journal}{Combust. Flame} \bibinfo{volume}{157}
  (\bibinfo{year}{2010}) \bibinfo{pages}{2333--2339}.

\bibitem[{Dooley et~al.(2012)Dooley, Won, Heyne, Farouk, Ju, Dryer, Kumar, Hui,
  Sung, Wang, Oehlschlaeger, Iyer, Iyer, Litzinger, Santoro, Malewicki, and
  Brezinsky}]{Dooley:2012bp}
\bibinfo{author}{S.~Dooley}, \bibinfo{author}{S.~H. Won},
  \bibinfo{author}{J.~Heyne}, \bibinfo{author}{T.~I. Farouk},
  \bibinfo{author}{Y.~Ju}, \bibinfo{author}{F.~L. Dryer},
  \bibinfo{author}{K.~Kumar}, \bibinfo{author}{X.~Hui}, \bibinfo{author}{C.~J.
  Sung}, \bibinfo{author}{H.~Wang}, \bibinfo{author}{M.~A. Oehlschlaeger},
  \bibinfo{author}{V.~Iyer}, \bibinfo{author}{S.~Iyer}, \bibinfo{author}{T.~A.
  Litzinger}, \bibinfo{author}{R.~J. Santoro}, \bibinfo{author}{T.~Malewicki},
  \bibinfo{author}{K.~Brezinsky}, \bibinfo{journal}{Combust. Flame}
  \bibinfo{volume}{159} (\bibinfo{year}{2012}) \bibinfo{pages}{1444--1466}.

\bibitem[{Violi et~al.(2002)Violi, Yan, Eddings, Sarofim, Granata, Faravelli,
  and Ranzi}]{Violi:2002hl}
\bibinfo{author}{A.~Violi}, \bibinfo{author}{S.~Yan}, \bibinfo{author}{E.~G.
  Eddings}, \bibinfo{author}{A.~F. Sarofim}, \bibinfo{author}{S.~Granata},
  \bibinfo{author}{T.~Faravelli}, \bibinfo{author}{E.~Ranzi},
  \bibinfo{journal}{Combust. Sci. Technol.} \bibinfo{volume}{174}
  (\bibinfo{year}{2002}) \bibinfo{pages}{399--417}.

\bibitem[{Lu and Law(2009)}]{Lu:2009gh}
\bibinfo{author}{T.~Lu}, \bibinfo{author}{C.~K. Law}, \bibinfo{journal}{Prog.
  Energy Comb. Sci.} \bibinfo{volume}{35} (\bibinfo{year}{2009})
  \bibinfo{pages}{192--215}.

\bibitem[{Westbrook et~al.(2009)Westbrook, Pitz, Herbinet, Curran, and
  Silke}]{Westbrook:2009}
\bibinfo{author}{C.~K. Westbrook}, \bibinfo{author}{W.~J. Pitz},
  \bibinfo{author}{O.~Herbinet}, \bibinfo{author}{H.~J. Curran},
  \bibinfo{author}{E.~J. Silke}, \bibinfo{journal}{Combust. Flame}
  \bibinfo{volume}{156} (\bibinfo{year}{2009}) \bibinfo{pages}{181--199}.

\bibitem[{Sarathy et~al.(2011)Sarathy, Westbrook, Mehl, Pitz, Togbe, Dagaut,
  Wang, Oehlschlaeger, Niemann, Seshadri, Veloo, Ji, Egolfopoulos, and
  Lu}]{Sarathy:2011}
\bibinfo{author}{S.~Sarathy}, \bibinfo{author}{C.~Westbrook},
  \bibinfo{author}{M.~Mehl}, \bibinfo{author}{W.~Pitz},
  \bibinfo{author}{C.~Togbe}, \bibinfo{author}{P.~Dagaut},
  \bibinfo{author}{H.~Wang}, \bibinfo{author}{M.~Oehlschlaeger},
  \bibinfo{author}{U.~Niemann}, \bibinfo{author}{K.~Seshadri},
  \bibinfo{author}{P.~Veloo}, \bibinfo{author}{C.~Ji},
  \bibinfo{author}{F.~Egolfopoulos}, \bibinfo{author}{T.~Lu},
  \bibinfo{journal}{Combustion and Flame} \bibinfo{volume}{158}
  (\bibinfo{year}{2011}) \bibinfo{pages}{2338--2357}.

\bibitem[{Blurock(1995)}]{Blurock:1995tj}
\bibinfo{author}{E.~S. Blurock}, \bibinfo{journal}{J. Chem. Inf. Comput. Sci.}
  \bibinfo{volume}{35} (\bibinfo{year}{1995}) \bibinfo{pages}{607--616}.

\bibitem[{Ranzi et~al.(1995)Ranzi, Faravelli, Gaffuri, and Sogaro}]{Ranzi:1995}
\bibinfo{author}{E.~Ranzi}, \bibinfo{author}{T.~Faravelli},
  \bibinfo{author}{P.~Gaffuri}, \bibinfo{author}{A.~Sogaro},
  \bibinfo{journal}{Combust. Flame} \bibinfo{volume}{102}
  (\bibinfo{year}{1995}) \bibinfo{pages}{179--192}.

\bibitem[{C{\^o}me et~al.(1996)C{\^o}me, Warth, Glaude, Fournet,
  Battin-Leclerc, and Scacchi}]{Come:1996}
\bibinfo{author}{G.~M. C{\^o}me}, \bibinfo{author}{V.~Warth},
  \bibinfo{author}{P.~A. Glaude}, \bibinfo{author}{R.~Fournet},
  \bibinfo{author}{F.~Battin-Leclerc}, \bibinfo{author}{G.~Scacchi},
  \bibinfo{journal}{Symp. (Int.) Combust.} \bibinfo{volume}{26}
  (\bibinfo{year}{1996}) \bibinfo{pages}{755--762}.

\bibitem[{Warth et~al.(2000)Warth, Battin-Leclerc, Fournet, Glaude, C{\^o}me,
  and Scacchi}]{Warth:2000bx}
\bibinfo{author}{V.~Warth}, \bibinfo{author}{F.~Battin-Leclerc},
  \bibinfo{author}{R.~Fournet}, \bibinfo{author}{P.~A. Glaude},
  \bibinfo{author}{G.~M. C{\^o}me}, \bibinfo{author}{G.~Scacchi},
  \bibinfo{journal}{Comput. Chem.} \bibinfo{volume}{24} (\bibinfo{year}{2000})
  \bibinfo{pages}{541--560}.

\bibitem[{Ranzi et~al.(2005)Ranzi, Frassoldati, Granata, and
  Faravelli}]{Ranzi:2005fy}
\bibinfo{author}{E.~Ranzi}, \bibinfo{author}{A.~Frassoldati},
  \bibinfo{author}{S.~Granata}, \bibinfo{author}{T.~Faravelli},
  \bibinfo{journal}{Ind. Eng. Chem. Res.} \bibinfo{volume}{44}
  (\bibinfo{year}{2005}) \bibinfo{pages}{5170--5183}.

\bibitem[{Mor{\'e}ac et~al.(2006)Mor{\'e}ac, Blurock, and
  Mauss}]{Moreac:2006ej}
\bibinfo{author}{G.~Mor{\'e}ac}, \bibinfo{author}{E.~S. Blurock},
  \bibinfo{author}{F.~Mauss}, \bibinfo{journal}{Combust. Sci. Technol.}
  \bibinfo{volume}{178} (\bibinfo{year}{2006}) \bibinfo{pages}{2025--2038}.

\bibitem[{Van~Geem et~al.(2006)Van~Geem, Reyniers, Marin, Song, Green, and
  Matheu}]{VanGeem:2006fv}
\bibinfo{author}{K.~M. Van~Geem}, \bibinfo{author}{M.-F. Reyniers},
  \bibinfo{author}{G.~B. Marin}, \bibinfo{author}{J.~Song},
  \bibinfo{author}{W.~H. Green}, \bibinfo{author}{D.~M. Matheu},
  \bibinfo{journal}{AICHe J.} \bibinfo{volume}{52} (\bibinfo{year}{2006})
  \bibinfo{pages}{718--730}.

\bibitem[{Hansen et~al.(2013)Hansen, Merchant, Harper, and
  Green}]{Hansen:2013fe}
\bibinfo{author}{N.~Hansen}, \bibinfo{author}{S.~S. Merchant},
  \bibinfo{author}{M.~R. Harper}, \bibinfo{author}{W.~H. Green},
  \bibinfo{journal}{Combust. Flame} \bibinfo{volume}{160}
  (\bibinfo{year}{2013}) \bibinfo{pages}{2343--2351}.

\bibitem[{Tomlin et~al.(1997)Tomlin, Tur{\'a}nyi, and Pilling}]{Tomlin:1997}
\bibinfo{author}{A.~S. Tomlin}, \bibinfo{author}{T.~Tur{\'a}nyi},
  \bibinfo{author}{M.~J. Pilling}, \bibinfo{journal}{Comprehensive Chemical
  Kinetics}  (\bibinfo{year}{1997}) \bibinfo{pages}{293--437}.

\bibitem[{Pierucci and Ranzi(2008)}]{Pierucci:2008dl}
\bibinfo{author}{S.~Pierucci}, \bibinfo{author}{E.~Ranzi},
  \bibinfo{journal}{Comput. Chem. Eng.} \bibinfo{volume}{32}
  (\bibinfo{year}{2008}) \bibinfo{pages}{805--826}.

\bibitem[{Blurock et~al.(2013)Blurock, Battin-Leclerc, Faravelli, and
  Green}]{Blurock:2013vc}
\bibinfo{author}{E.~Blurock}, \bibinfo{author}{F.~Battin-Leclerc},
  \bibinfo{author}{T.~Faravelli}, \bibinfo{author}{W.~H. Green}, in:
  \bibinfo{editor}{F.~Battin-Leclerc}, \bibinfo{editor}{J.~M. Simmie},
  \bibinfo{editor}{E.~Blurock} (Eds.), \bibinfo{booktitle}{Cleaner Combustion},
  Green Energy and Technology, Springer-Verlag, \bibinfo{address}{London},
  \bibinfo{year}{2013}, pp. \bibinfo{pages}{59--92}.

\bibitem[{Lu and Law(2005)}]{Lu:2005ce}
\bibinfo{author}{T.~Lu}, \bibinfo{author}{C.~K. Law}, \bibinfo{journal}{Proc.
  Combust. Inst.} \bibinfo{volume}{30} (\bibinfo{year}{2005})
  \bibinfo{pages}{1333--1341}.

\bibitem[{Lu and Law(2006{\natexlab{a}})}]{Lu:2006bb}
\bibinfo{author}{T.~Lu}, \bibinfo{author}{C.~K. Law},
  \bibinfo{journal}{Combust. Flame} \bibinfo{volume}{144}
  (\bibinfo{year}{2006}{\natexlab{a}}) \bibinfo{pages}{24--36}.

\bibitem[{Lu and Law(2006{\natexlab{b}})}]{Lu:2006gi}
\bibinfo{author}{T.~Lu}, \bibinfo{author}{C.~K. Law},
  \bibinfo{journal}{Combust. Flame} \bibinfo{volume}{146}
  (\bibinfo{year}{2006}{\natexlab{b}}) \bibinfo{pages}{472--483}.

\bibitem[{Bendtsen et~al.(2001)Bendtsen, Glarborg, and
  Dam-Johansen}]{Bendtsen:2001vh}
\bibinfo{author}{A.~Bendtsen}, \bibinfo{author}{P.~Glarborg},
  \bibinfo{author}{K.~Dam-Johansen}, \bibinfo{journal}{Comput. Chem.}
  \bibinfo{volume}{25} (\bibinfo{year}{2001}) \bibinfo{pages}{161--170}.

\bibitem[{Pepiot-Desjardins and Pitsch(2008)}]{Pepiot-Desjardins:2008}
\bibinfo{author}{P.~Pepiot-Desjardins}, \bibinfo{author}{H.~Pitsch},
  \bibinfo{journal}{Combust. Flame} \bibinfo{volume}{154}
  (\bibinfo{year}{2008}) \bibinfo{pages}{67--81}.

\bibitem[{Niemeyer and Sung(2011)}]{Niemeyer:2011fe}
\bibinfo{author}{K.~E. Niemeyer}, \bibinfo{author}{C.~J. Sung},
  \bibinfo{journal}{Combust. Flame} \bibinfo{volume}{158}
  (\bibinfo{year}{2011}) \bibinfo{pages}{1439--1443}.

\bibitem[{Sun et~al.(2010)Sun, Chen, Gou, and Ju}]{Sun:2010jf}
\bibinfo{author}{W.~Sun}, \bibinfo{author}{Z.~Chen}, \bibinfo{author}{X.~Gou},
  \bibinfo{author}{Y.~Ju}, \bibinfo{journal}{Combust. Flame}
  \bibinfo{volume}{157} (\bibinfo{year}{2010}) \bibinfo{pages}{1298--1307}.

\bibitem[{Sankaran et~al.(2007)Sankaran, Hawkes, Chen, Lu, and
  Law}]{Sankaran:2007fs}
\bibinfo{author}{R.~Sankaran}, \bibinfo{author}{E.~R. Hawkes},
  \bibinfo{author}{J.~H. Chen}, \bibinfo{author}{T.~Lu}, \bibinfo{author}{C.~K.
  Law}, \bibinfo{journal}{Proc. Combust. Inst.} \bibinfo{volume}{31}
  (\bibinfo{year}{2007}) \bibinfo{pages}{1291--1298}.

\bibitem[{Zheng et~al.(2007)Zheng, Lu, and Law}]{Zheng:2007gd}
\bibinfo{author}{X.~L. Zheng}, \bibinfo{author}{T.~Lu}, \bibinfo{author}{C.~K.
  Law}, \bibinfo{journal}{Proc. Combust. Inst.} \bibinfo{volume}{31}
  (\bibinfo{year}{2007}) \bibinfo{pages}{367--375}.

\bibitem[{Niemeyer et~al.(2010)Niemeyer, Sung, and Raju}]{Niemeyer:2010bt}
\bibinfo{author}{K.~E. Niemeyer}, \bibinfo{author}{C.~J. Sung},
  \bibinfo{author}{M.~P. Raju}, \bibinfo{journal}{Combust. Flame}
  \bibinfo{volume}{157} (\bibinfo{year}{2010}) \bibinfo{pages}{1760--1770}.

\bibitem[{Luo et~al.(2010)Luo, Lu, Maciaszek, Som, and Longman}]{Luo:2010cu}
\bibinfo{author}{Z.~Luo}, \bibinfo{author}{T.~Lu}, \bibinfo{author}{M.~J.
  Maciaszek}, \bibinfo{author}{S.~K. Som}, \bibinfo{author}{D.~E. Longman},
  \bibinfo{journal}{Energy Fuels} \bibinfo{volume}{24} (\bibinfo{year}{2010})
  \bibinfo{pages}{6283--6293}.

\bibitem[{Shi et~al.(2010)Shi, Ge, Brakora, and Reitz}]{Shi:2010a}
\bibinfo{author}{Y.~Shi}, \bibinfo{author}{H.-W. Ge}, \bibinfo{author}{J.~L.
  Brakora}, \bibinfo{author}{R.~D. Reitz}, \bibinfo{journal}{Energy Fuels}
  \bibinfo{volume}{24} (\bibinfo{year}{2010}) \bibinfo{pages}{1646--1654}.

\bibitem[{Bahlouli et~al.(2012)Bahlouli, Saray, and Atikol}]{Bahlouli:2012kq}
\bibinfo{author}{K.~Bahlouli}, \bibinfo{author}{R.~K. Saray},
  \bibinfo{author}{U.~Atikol}, \bibinfo{journal}{Energy Fuels}
  \bibinfo{volume}{26} (\bibinfo{year}{2012}) \bibinfo{pages}{3244---3256}.

\bibitem[{Shi et~al.(2010)Shi, Liang, Ge, and Reitz}]{Shi:2010}
\bibinfo{author}{Y.~Shi}, \bibinfo{author}{L.~Liang}, \bibinfo{author}{H.-W.
  Ge}, \bibinfo{author}{R.~D. Reitz}, \bibinfo{journal}{Combust. Theor. Model.}
  \bibinfo{volume}{14} (\bibinfo{year}{2010}) \bibinfo{pages}{69--89}.

\bibitem[{Luo et~al.(2012{\natexlab{a}})Luo, Plomer, Lu, Som, Longman, Sarathy,
  and Pitz}]{Luo:2012dp}
\bibinfo{author}{Z.~Luo}, \bibinfo{author}{M.~Plomer}, \bibinfo{author}{T.~Lu},
  \bibinfo{author}{S.~K. Som}, \bibinfo{author}{D.~E. Longman},
  \bibinfo{author}{S.~M. Sarathy}, \bibinfo{author}{W.~J. Pitz},
  \bibinfo{journal}{Fuel} \bibinfo{volume}{99}
  (\bibinfo{year}{2012}{\natexlab{a}}) \bibinfo{pages}{143--153}.

\bibitem[{Luo et~al.(2012{\natexlab{b}})Luo, Plomer, Lu, Som, and
  Longman}]{Luo:2012cr}
\bibinfo{author}{Z.~Luo}, \bibinfo{author}{M.~Plomer}, \bibinfo{author}{T.~Lu},
  \bibinfo{author}{S.~K. Som}, \bibinfo{author}{D.~E. Longman},
  \bibinfo{journal}{Combust. Theor. Model.} \bibinfo{volume}{16}
  (\bibinfo{year}{2012}{\natexlab{b}}) \bibinfo{pages}{369--385}.

\bibitem[{Blanquart et~al.(2009)Blanquart, Pepiot-Desjardins, and
  Pitsch}]{Blanquart:2009iw}
\bibinfo{author}{G.~Blanquart}, \bibinfo{author}{P.~Pepiot-Desjardins},
  \bibinfo{author}{H.~Pitsch}, \bibinfo{journal}{Combust. Flame}
  \bibinfo{volume}{156} (\bibinfo{year}{2009}) \bibinfo{pages}{588--607}.

\bibitem[{Ra and Reitz(2011)}]{Ra:2011gh}
\bibinfo{author}{Y.~Ra}, \bibinfo{author}{R.~D. Reitz},
  \bibinfo{journal}{Combust. Flame} \bibinfo{volume}{158}
  (\bibinfo{year}{2011}) \bibinfo{pages}{69--90}.

\bibitem[{Lu et~al.(2001)Lu, Ju, and Law}]{Lu:2001ve}
\bibinfo{author}{T.~Lu}, \bibinfo{author}{Y.~Ju}, \bibinfo{author}{C.~K. Law},
  \bibinfo{journal}{Combust. Flame} \bibinfo{volume}{126}
  (\bibinfo{year}{2001}) \bibinfo{pages}{1445--1455}.

\bibitem[{Niemeyer(2010)}]{Niemeyer:2010}
\bibinfo{author}{K.~E. Niemeyer}, \bibinfo{title}{Skeletal Mechanism Generation
  for Surrogate Fuels}, \bibinfo{type}{{MS} thesis}, Case Western Reserve
  University, \bibinfo{year}{2010}.

\bibitem[{Lutz et~al.(1997)Lutz, Kee, and Miller}]{Lutz:1997vu}
\bibinfo{author}{A.~E. Lutz}, \bibinfo{author}{R.~J. Kee},
  \bibinfo{author}{J.~A. Miller}, \bibinfo{title}{{SENKIN}: A {FORTRAN} Program
  for Predicting Homogeneous Gas Phase Chemical Kinetics with Sensitivity
  Analysis}, \bibinfo{type}{Technical Report} \bibinfo{number}{SAND87-8248},
  Sandia National Laboratories, \bibinfo{year}{1997}.

\bibitem[{Kee et~al.(1996)Kee, Rupley, Meeks, and Miller}]{Kee:1996vd}
\bibinfo{author}{R.~J. Kee}, \bibinfo{author}{F.~M. Rupley},
  \bibinfo{author}{E.~Meeks}, \bibinfo{author}{J.~A. Miller},
  \bibinfo{title}{{CHEMKIN-III}: A {FORTRAN} Chemical Kinetics Package for the
  Analysis of Gas-Phase Chemical and Plasma Kinetics}, \bibinfo{type}{Technical
  Report} \bibinfo{number}{SAND96-8216}, Sandia National Laboratories,
  \bibinfo{year}{1996}.

\bibitem[{Chaos et~al.(2007)Chaos, Kazakov, Zhao, and Dryer}]{Chaos:2007fc}
\bibinfo{author}{M.~Chaos}, \bibinfo{author}{A.~Kazakov},
  \bibinfo{author}{Z.~Zhao}, \bibinfo{author}{F.~L. Dryer},
  \bibinfo{journal}{Int. J. Chem. Kinet.} \bibinfo{volume}{39}
  (\bibinfo{year}{2007}) \bibinfo{pages}{399--414}.

\bibitem[{Hansen et~al.(2011)Hansen, Harper, and Green}]{Hansen:2011bu}
\bibinfo{author}{N.~Hansen}, \bibinfo{author}{M.~R. Harper},
  \bibinfo{author}{W.~H. Green}, \bibinfo{journal}{Phys. Chem. Chem. Phys.}
  \bibinfo{volume}{13} (\bibinfo{year}{2011}) \bibinfo{pages}{20262--20274}.

\bibitem[{Blurock and Battin-Leclerc(2013)}]{Blurock:2013}
\bibinfo{author}{E.~Blurock}, \bibinfo{author}{F.~Battin-Leclerc}, in:
  \bibinfo{editor}{F.~Battin-Leclerc}, \bibinfo{editor}{J.~M. Simmie},
  \bibinfo{editor}{E.~Blurock} (Eds.), \bibinfo{booktitle}{Cleaner Combustion},
  Green Energy and Technology, Springer-Verlag, \bibinfo{address}{London},
  \bibinfo{year}{2013}, pp. \bibinfo{pages}{17--58}.

\bibitem[{Miller et~al.(1990)Miller, Kee, and Westbrook}]{Miller:1990et}
\bibinfo{author}{J.~A. Miller}, \bibinfo{author}{R.~J. Kee},
  \bibinfo{author}{C.~K. Westbrook}, \bibinfo{journal}{Annu. Rev. Phys. Chem.}
  \bibinfo{volume}{41} (\bibinfo{year}{1990}) \bibinfo{pages}{345--387}.

\bibitem[{Richter and Howard(2000)}]{Richter:2000uy}
\bibinfo{author}{H.~Richter}, \bibinfo{author}{J.~B. Howard},
  \bibinfo{journal}{Prog. Energy Comb. Sci.} \bibinfo{volume}{26}
  (\bibinfo{year}{2000}) \bibinfo{pages}{565--608}.

\bibitem[{Hansen et~al.(2006)Hansen, Klippenstein, Taatjes, Miller, Wang, Cool,
  Yang, Yang, Wei, Huang, Wang, Qi, Law, and Westmoreland}]{Hansen:2006hj}
\bibinfo{author}{N.~Hansen}, \bibinfo{author}{S.~J. Klippenstein},
  \bibinfo{author}{C.~A. Taatjes}, \bibinfo{author}{J.~A. Miller},
  \bibinfo{author}{J.~Wang}, \bibinfo{author}{T.~A. Cool},
  \bibinfo{author}{B.~Yang}, \bibinfo{author}{R.~Yang},
  \bibinfo{author}{L.~Wei}, \bibinfo{author}{C.~Huang},
  \bibinfo{author}{J.~Wang}, \bibinfo{author}{F.~Qi}, \bibinfo{author}{M.~E.
  Law}, \bibinfo{author}{P.~R. Westmoreland}, \bibinfo{journal}{J. Phys. Chem.
  A} \bibinfo{volume}{110} (\bibinfo{year}{2006}) \bibinfo{pages}{3670--3678}.

\bibitem[{Lories et~al.(2010)Lories, Vandooren, and Peeters}]{Lories:2010gp}
\bibinfo{author}{X.~Lories}, \bibinfo{author}{J.~Vandooren},
  \bibinfo{author}{D.~Peeters}, \bibinfo{journal}{Phys. Chem. Chem. Phys.}
  \bibinfo{volume}{12} (\bibinfo{year}{2010}) \bibinfo{pages}{3762}.

\bibitem[{Valorani et~al.(2006)Valorani, Creta, Goussis, Lee, and
  Najm}]{Valorani:2006bp}
\bibinfo{author}{M.~Valorani}, \bibinfo{author}{F.~Creta},
  \bibinfo{author}{D.~A. Goussis}, \bibinfo{author}{J.~C. Lee},
  \bibinfo{author}{H.~N. Najm}, \bibinfo{journal}{Combust. Flame}
  \bibinfo{volume}{146} (\bibinfo{year}{2006}) \bibinfo{pages}{29--51}.

\bibitem[{Valorani et~al.(2007)Valorani, Creta, Donato, Najm, and
  Goussis}]{Valorani:2007ef}
\bibinfo{author}{M.~Valorani}, \bibinfo{author}{F.~Creta},
  \bibinfo{author}{F.~Donato}, \bibinfo{author}{H.~N. Najm},
  \bibinfo{author}{D.~A. Goussis}, \bibinfo{journal}{Proc. Combust. Inst.}
  \bibinfo{volume}{31} (\bibinfo{year}{2007}) \bibinfo{pages}{483--490}.

\bibitem[{Mitsos et~al.(2008)Mitsos, Oxberry, Barton, and Green}]{Mitsos:2008}
\bibinfo{author}{A.~Mitsos}, \bibinfo{author}{G.~Oxberry},
  \bibinfo{author}{P.~I. Barton}, \bibinfo{author}{W.~Green},
  \bibinfo{journal}{Combust. Flame} \bibinfo{volume}{155}
  (\bibinfo{year}{2008}) \bibinfo{pages}{118--132}.

\bibitem[{Prager et~al.(2009)Prager, Najm, Valorani, and
  Goussis}]{Prager:2009ge}
\bibinfo{author}{J.~Prager}, \bibinfo{author}{H.~N. Najm},
  \bibinfo{author}{M.~Valorani}, \bibinfo{author}{D.~A. Goussis},
  \bibinfo{journal}{Proc. Combust. Inst.} \bibinfo{volume}{32}
  (\bibinfo{year}{2009}) \bibinfo{pages}{509--517}.

\bibitem[{Zs{\'e}ly et~al.(2011)Zs{\'e}ly, Nagy, Simmie, and
  Curran}]{Zsely:2011im}
\bibinfo{author}{I.~G. Zs{\'e}ly}, \bibinfo{author}{T.~Nagy},
  \bibinfo{author}{J.~M. Simmie}, \bibinfo{author}{H.~J. Curran},
  \bibinfo{journal}{Combust. Flame} \bibinfo{volume}{158}
  (\bibinfo{year}{2011}) \bibinfo{pages}{1469--1479}.

\bibitem[{{Reaction Design: San Diego}(2012)}]{chemkin:2012}
\bibinfo{author}{{Reaction Design: San Diego}}, \bibinfo{title}{{CHEMKIN-PRO}
  15113}, \bibinfo{year}{2012}.

\bibitem[{Sakai et~al.(2009)Sakai, Miyoshi, Koshi, and Pitz}]{Sakai:2009bp}
\bibinfo{author}{Y.~Sakai}, \bibinfo{author}{A.~Miyoshi},
  \bibinfo{author}{M.~Koshi}, \bibinfo{author}{W.~J. Pitz},
  \bibinfo{journal}{Proc. Combust. Inst.} \bibinfo{volume}{32}
  (\bibinfo{year}{2009}) \bibinfo{pages}{411--418}.

\bibitem[{Battin-Leclerc(2008)}]{Battin-Leclerc:2008}
\bibinfo{author}{F.~Battin-Leclerc}, \bibinfo{journal}{Prog. Energy Comb. Sci.}
  \bibinfo{volume}{34} (\bibinfo{year}{2008}) \bibinfo{pages}{440--498}.

\bibitem[{Z{\'a}dor et~al.(2011)Z{\'a}dor, Taatjes, and
  Fernandes}]{Zador:2011kz}
\bibinfo{author}{J.~Z{\'a}dor}, \bibinfo{author}{C.~A. Taatjes},
  \bibinfo{author}{R.~X. Fernandes}, \bibinfo{journal}{Prog. Energy Comb. Sci.}
  \bibinfo{volume}{37} (\bibinfo{year}{2011}) \bibinfo{pages}{371--421}.

\bibitem[{Sj{\"o}berg et~al.(2007)Sj{\"o}berg, Dec, and Hwang}]{Sjoberg:2007}
\bibinfo{author}{M.~Sj{\"o}berg}, \bibinfo{author}{J.~E. Dec},
  \bibinfo{author}{W.~Hwang}, \bibinfo{title}{Thermodynamic and chemical
  effects of {EGR} and its constituents on {HCCI} autoignition},
  \bibinfo{howpublished}{{SAE} 2007-01-0207}, \bibinfo{year}{2007}.

\bibitem[{Liang et~al.(2009)Liang, Stevens, and Farrell}]{Liang:2009}
\bibinfo{author}{L.~Liang}, \bibinfo{author}{J.~Stevens},
  \bibinfo{author}{J.~T. Farrell}, \bibinfo{journal}{Proc. Combust. Inst.}
  \bibinfo{volume}{32} (\bibinfo{year}{2009}) \bibinfo{pages}{527--534}.

\bibitem[{Gou et~al.(2013)Gou, Chen, Sun, and Ju}]{Gou:2013eu}
\bibinfo{author}{X.~Gou}, \bibinfo{author}{Z.~Chen}, \bibinfo{author}{W.~Sun},
  \bibinfo{author}{Y.~Ju}, \bibinfo{journal}{Combust. Flame}
  \bibinfo{volume}{160} (\bibinfo{year}{2013}) \bibinfo{pages}{225--231}.

\bibitem[{Yang et~al.(2013)Yang, Ren, Lu, and Goldin}]{Yang:2013ip}
\bibinfo{author}{H.~Yang}, \bibinfo{author}{Z.~Ren}, \bibinfo{author}{T.~Lu},
  \bibinfo{author}{G.~M. Goldin}, \bibinfo{journal}{Combust. Theor. Model.}
  \bibinfo{volume}{17} (\bibinfo{year}{2013}) \bibinfo{pages}{167--183}.

\end{thebibliography}
\end{document}